\documentclass[final,5p,10pt]{elsarticle}

\usepackage{float}
\usepackage{epstopdf}
\usepackage[usenames, dvipsnames]{color}
\usepackage{soul}
\usepackage{multirow}

\usepackage{graphicx}
\usepackage{amssymb}
\usepackage{amsthm}
\usepackage{graphicx}
\usepackage[breaklinks=true,colorlinks=true,linkcolor=blue,urlcolor=blue,citecolor=blue]{hyperref}
\usepackage{framed}
\expandafter\let\csname equation*\endcsname\relax
\expandafter\let\csname endequation*\endcsname\relax
\usepackage{amsmath}
\usepackage{mathtools}
\usepackage{subcaption}
\usepackage{marginnote}
\reversemarginpar

\newcounter{para}

\usepackage{natbib}
\usepackage[utf8]{inputenc}
\usepackage{ascii}
\usepackage{newunicodechar}
\usepackage{gensymb} 

\biboptions{sort&compress}
\makeatletter
\def\ps@pprintTitle{%
	\let\@oddhead\@empty
	\let\@evenhead\@empty
	\def\@oddfoot{}%
	\let\@evenfoot\@oddfoot}
\makeatother
\journal{Journal Name}

\usepackage{epsfig}
\usepackage{euscript}
\usepackage{amsfonts}
\usepackage{url}

\def\ket#1{{|#1\rangle}}
\def\bra#1{{\langle#1|}}

\def\cg(#1,#2)(#3,#4)(#5,#6){\bra{#1,#2,#3,#4}#5,#6\rangle}

\def\ts#1{{_{\mbox{\scriptsize #1}}}}

\def\threej(#1,#2)(#3,#4)(#5,#6){\begin{pmatrix}#1&#3&#5\\#2&#4&#6\end{pmatrix}}
\def\sixj(#1,#2,#3)(#4,#5,#6){\begin{Bmatrix}#1&#2&#3\\#4&#5&#6\end{Bmatrix}}
\def\ninej(#1,#2,#3)(#4,#5,#6)(#7,#8,#9){\begin{Bmatrix}#1&#2&#3\\#4&#5&#6\\#7&#8&#9\end{Bmatrix}}

\def\bs{\boldsymbol}

\newlength{\defbaselineskip}
\newcommand{\setlinespacing}[1]%
{\setlength{\baselineskip}{#1 \defbaselineskip}}

\usepackage{mathtools, cuted}
\begin{document}
	\begin{frontmatter}
		\title{Constraints on bosonic dark matter from ultralow-field nuclear magnetic resonance
		}
		
		\author[1,2]{Antoine~Garcon}
		\author[2]{John~W.~Blanchard\corref{cor1}%
		}
		\author[1,2]{Gary P. Centers}
		\author[1,2]{Nataniel~L.~Figueroa}
		\author[3]{Peter~W.~Graham}
		\author[4]{Derek~F.~Jackson~Kimball}
		\author[5]{Surjeet~Rajendran}
		\author[6]{Alexander~O.~Sushkov}
		\author[1,2]{Yevgeny~V.~Stadnik}
		\author[1,2]{Arne~Wickenbrock}
		\author[1,2]{Teng~Wu}
		\author[1,2,5,7]{Dmitry~Budker}

		\address[1]{Johannes Gutenberg-Universit{\"a}t, Mainz 55128, Germany}
		\address[2]{Helmholtz-Institut Mainz, 55128 Mainz, Germany}
		\address[3]{Department of Physics, Stanford Institute for Theoretical Physics, Stanford University, California 94305, USA}
		\address[4]{Department of Physics, California State University East Bay, Hayward, California 94542-3084,USA}
		\address[5]{Department of Physics, University of California, Berkeley, CA 94720-7300,USA}
		\address[6]{Department of Physics, Boston University, Boston, Massachusetts 02215, USA}
		\address[7]{Nuclear Science Division, Lawrence Berkeley National Laboratory, Berkeley, CA 94720,USA}
		
		\cortext[cor1]{Corresponding author: blanchard@uni-mainz.de}


		\begin{abstract}
The nature of dark matter, the invisible substance making up over $80\%$ of the matter in the Universe, is one of the most fundamental mysteries of modern physics. 
Ultralight bosons such as axions, axion-like particles or dark photons could make up most of the dark matter.
Couplings between such bosons and nuclear spins may enable their direct detection via nuclear magnetic resonance (NMR) spectroscopy: as nuclear spins move through the galactic dark-matter halo, they couple to dark-matter and behave as if they were in an oscillating magnetic field, generating a dark-matter-driven NMR signal.
As part of the Cosmic Axion Spin Precession Experiment (CASPEr), an NMR-based dark-matter search, we use ultralow-field NMR to probe the axion-fermion ``wind" coupling and dark-photon couplings to nuclear spins.
No dark matter signal was detected above background, establishing new experimental bounds for dark-matter bosons with masses ranging from \mbox{$1.8\times 10^{-16}$ to $7.8\times 10^{-14}$ eV.}
		\end{abstract}
	\end{frontmatter}
	\section{\textbf{INTRODUCTION}}\label{s:10}
	\subsection{\textbf{Ultralight bosonic dark matter}}\label{ss:11}
The nature of dark matter, the invisible substance that makes up over $80\%$ of the matter in the universe \cite{Ackerman2009}, is one of the most intriguing mysteries of modern physics \cite{Bertone2018}. 
Elucidating the nature of dark matter will profoundly impact our understanding of cosmology, astrophysics, and particle physics, providing insights into the evolution of the Universe and potentially uncovering new physical laws and fundamental forces beyond the Standard Model. 
While the observational evidence for dark matter is derived from its gravitational effects at the galactic scale and larger, the key to solving the mystery of its nature lies in directly measuring non-gravitational interactions of dark matter with Standard Model particles and fields.
	
To date, experimental efforts to directly detect dark matter have largely focused on Weakly Interacting Massive Particles (WIMPs), with masses between 10 and 1000 GeV \cite{Ber05,Fen10}. 
Despite considerable efforts, there have been no conclusive signs of WIMP interactions with ordinary matter. 
The absence of evidence for WIMPs has reinvigorated efforts to search for ultralight bosonic fields, another class of theoretically well-motivated dark matter candidates \cite{Gra15review}, composed of bosons with masses smaller than a few eV. 
A wide variety of theories predict new spin-0 bosons such as axions and axion-like particles (ALPs) as well as spin-1 bosons such as dark photons \cite{Ackerman2009,Saf17review}. 
Their existence may help to answer other open questions in particle physics such as why the strong force respects the combined charge-conjugation and parity-inversion (CP) symmetry to such a high degree \cite{Pec77}, the relative weakness of the gravitational interaction \cite{Gra15}, and how to unify the theories of quantum mechanics and general relativity \cite{Svr06}.
	
Bosonic fields can be detected through their interactions with Standard Model particles. 
Most experiments searching for bosonic fields seek to detect photons created by the conversion of these bosons in strong electromagnetic fields via the Primakoff effect  \cite{ADMX1,CAST1,ALP1,Bra03,Bru16,Sem17,ADMX2}. 
Another method to search for dark-matter bosonic fields was recently proposed: dark-matter-driven spin-precession \cite{nedm13,Bud14,Gra13}, detected via nuclear magnetic resonance (NMR) techniques \cite{Bud14,Gra13,Kim17,Wan17,nEdmFlambaum}. These concepts were recently applied to data measuring the permanent electric dipole moment (EDM) of the neutron and successfully constrained ALP dark matter with masses $\lesssim 10^{-17}$ eV \cite{nEdmFlambaum}.
	
The Cosmic Axion Spin Precession Experiment (CASPEr) is a multi-faceted research program using NMR techniques to search for dark-matter-driven spin precession \cite{Bud14}. 
Efforts within this program utilizing zero- to ultralow-field (ZULF) NMR spectroscopy \cite{Blanchard2016d} are collectively referred to as CASPEr-ZULF.
We report here experimental results from a CASPER-ZULF search for ultralight bosonic dark-matter fields, probing bosons with masses ranging from \mbox{$1.8\times 10^{-16}$} to \mbox{$7.8\times 10^{-14}$~eV} (corresponding to Compton frequencies ranging from \mbox{$\sim 45$~mHz} to $19$~Hz).
	
In the following section, we explain how dark-matter bosonic fields, in particular axion, ALP and dark-photon fields, can be detected by examining ZULF NMR spectra. 
The subsequent sections describe the measurement scheme and the data processing techniques employed during this search. 
Finally we present new laboratory results for bosonic dark matter, complementing astrophysical constraints obtained from supernovae $1987A$\footnote{We note that the constraints based on  SN$1987$A data continue to be reexamined; see e.g. Refs.~\cite{Blum2016,SNcontest}.}~\cite{Vysotsky}. 
	
	\section{\textbf{RESULTS}}\label{s:20}
	
	\subsection{\textbf{Dark-matter field properties}}
	
	If dark matter predominantly consists of particles with masses $m\ts{DM} \lesssim 10~{\rm eV}$, making up the totality of the average local dark-matter density, then they must be bosons with a large mode occupation number. 
	It would be impossible for fermions with such low masses to account for the observed galactic dark matter density, since the Pauli exclusion principle prevents them from having the required mode occupation. 
	
	In this scenario, axion and ALP bosonic dark matter is well described by a classical field $a(t)$, oscillating at the Compton frequency \mbox{($\omega\ts{DM}\approx m\ts{DM}c^2/\hbar$)} \cite{Pre83,Din83,ABBOTT1983133}:
	\begin{align}
	a(t) \approx a_0 \cos(\omega\ts{DM} t)~,
	\label{Eq:oscillating-axion-field}
	\end{align}
	where $c$ is the velocity of light, $\hbar$ is the reduced Planck constant and $a_0$ is the amplitude of the bosonic field.
	
	The temporal coherence of the bosonic field is limited by the relative motion of the Earth through random spatial fluctuations of the field. The characteristic coherence time $\tau\ts{DM}$, during which the bosonic dark-matter fields remains phase coherent, corresponds to $\sim 10^6$ periods of oscillation of the fields \cite{Bud14}.
	
	The amplitude $a_0$ can be estimated by assuming that the field energy density constitutes the totality of the average local dark-matter energy density \mbox{($\rho\ts{DM} \approx 0.4~{\rm GeV/cm^3}$ \cite{DMdensity}).} Then $a_0$ is related to the dark-matter density through:
	\begin{equation}
	\rho\ts{DM} = \frac{1}{2 }\frac{c^2}{\hbar^2} m\ts{DM} ^2  a_0 ^2~.
	\label{Eq:a0-from-DM-density}
	\end{equation}
	We note that $a_0$ is expected to fluctuate with relative amplitude of order one due to self-interference of the field.
	For simplicity we assume  \mbox{$a_0$} to be constant.
	\begin{figure*} 
		\centering
		\includegraphics{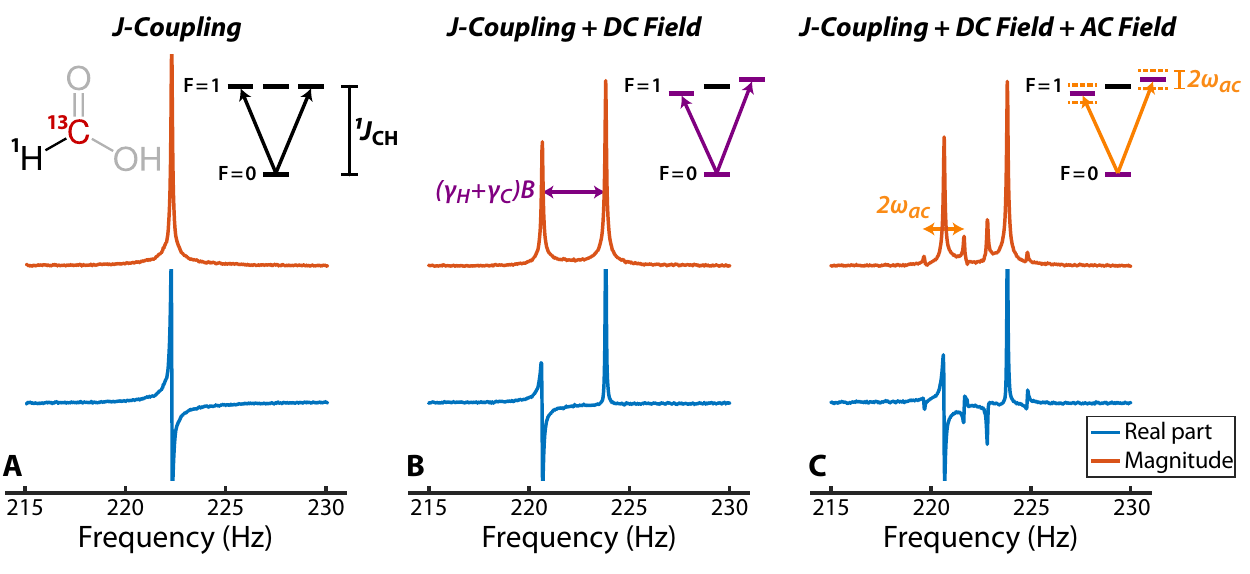}
		\caption{Nuclear spin energy levels and NMR spectra of $^{13}$C-formic acid measured in three different field conditions. \textbf{(a) } At zero magnetic field, the $F=1$ levels are degenerate, resulting in a spectrum exhibiting a single peak at the $J$-coupling frequency. \textbf{(b)} In the presence of a DC-magnetic field $B\ts{z} \approx 50$ nT, the $m_F=\pm 1$ degeneracy is lifted. The spectrum exhibits two split $J$-resonances. The splitting is equal to $\hbar m_F B\ts{z} (\gamma_\text{C}  + \gamma_\text{H})$.  The asymmetry of the resonances is due  to the influence of the applied field on the response characteristics of the atomic magnetometer. \textbf{(c)} The addition of an oscillating magnetic field along $B\ts{z}$ modulates the $m_F=\pm 1$ energy levels, resulting in sidebands located at $J/2\pi  \pm B\ts{z}(\gamma_\text{C}  + \gamma_\text{H}) / 2 \pm \omega_{AC}$ with amplitude proportional to the modulation index: $A\ts{s}\propto B\ts{AC}(\gamma_\text{C}  + \gamma_\text{H})  / 2 \omega\ts{AC}$.}
		\label{fig:formicacid}
	\end{figure*}
	
	\subsection{\textbf{Dark-matter couplings to nuclear spins}}\label{ss:12}
	
	CASPEr-Wind is sensitive to any field such that its interaction with nuclear spins can be written in the form:
	\begin{equation}
	\mathcal{H}\ts{DM} = - \hbar g\ts{N} \bs{I}\ts{N} \cdot \bs{D},
	\label{Eq:wind-Hamiltonian}
	\end{equation}
	where $g\ts{N}$ is a coupling constant that parametrizes the coupling of the effective field, $\bs{D}$, to nuclear spins represented by the operator $\bs{I}\ts{N}$.
	In analogy with the Zeeman interaction, \mbox{$\mathcal{H}\ts{Zeeman} = - \hbar \gamma\ts{N} \bs{I}\ts{N} \cdot \bs{B}$}, such an effective field $\bs{D}$ may be thought of as a pseudo-magnetic field interacting with nuclear spins, where the nuclear gyromagnetic ratio, $\gamma\ts{N}$, is replaced by the coupling constant, $g\ts{N}$.
	
	For clarity, we focus this discussion on the the so-called ``axion wind interaction'' with effective field \mbox{$\bs{D}\ts{wind}=-\bs{\nabla} a(\bs{r},t)$} and coupling constant $g\ts{aNN}$.
	A number of other possible couplings between nuclear spins and bosonic dark matter fields take a form similar to Eq.~\eqref{Eq:wind-Hamiltonian}.
	These include couplings to the ``dark'' electric (with coupling constant $g\ts{dEDM}$ and effective field $\bs{D}\ts{dEDM}$) and magnetic (with coupling constant $g\ts{dMDM}$ and effective field $\bs{D}\ts{dMDM}$) fields mediated by spin-$1$ bosons such as dark photons \cite{Gra15review,PhysRevD.84.103501} or a quadratic ``wind'' coupling to an ALP field (with coupling constant $g\ts{quad}$ and effective field $\bs{D}\ts{quad}$) \cite{Pos13}. 
	These are discussed further in the Materials and Methods section. 

	As the Earth orbits around the Sun (itself moving towards the Cygnus constellation at velocity, $\bs{v}$, comparable to the local galactic virial velocity $\sim~10^{-3}c$), it moves through the galactic dark-matter halo and an interaction between axions and ALPs with a given nucleon $N$, can arise.
	Assuming that ALPs make up all of the dark matter energy density, $\rho\ts{DM}$, and that the dominant interaction with nucleon spins is linear in $\bs{\nabla} a(\bs{r},t)$, Eqs.~\eqref{Eq:oscillating-axion-field} and \eqref{Eq:a0-from-DM-density} can be used to write the the effective field as
	\begin{equation}
	\bs{D}\ts{wind}(t)= - \sqrt{2\hbar c \rho_{\rm{DM}}} \sin \left(\omega_{\rm{DM}} t \right) \bs{v}.
	\label{Eq:DM-D-Operator}
	\end{equation}
	
	Then, given the local galactic virial velocity, $\bs{v}$, only two free parameters remain in the Hamiltonian in Eq.~\eqref{Eq:wind-Hamiltonian}: the coupling constant, $g\ts{aNN}$, and the field's oscillation frequency, $\omega\ts{DM}$, fixed by the boson mass.
	A CASPEr search consists of probing this parameter space over a bosonic mass range defined by the bandwidth of the experiment.	
	In order to calibrate the experiment, we apply known magnetic fields and the experiment's sensitivity to magnetic fields directly translates to sensitivity to the coupling constant. 
	If no ALP field is detected, upper bounds on the coupling constant can be determined based on the overall sensitivity of the experiment.
	
	\begin{figure*}
		\centering
		\includegraphics{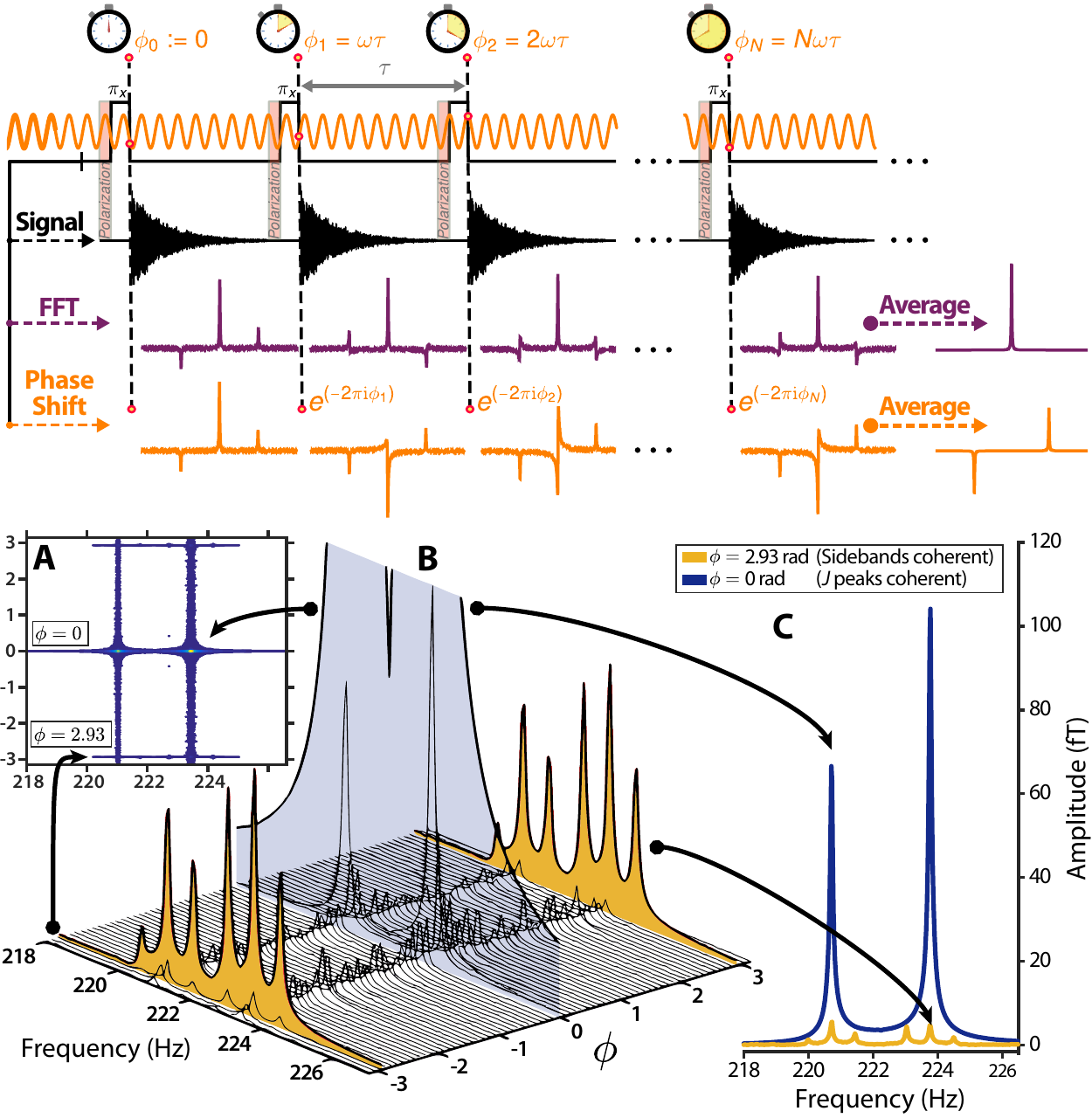}
		\caption{\textbf{Top}: signal acquisition scheme with simulated spectra. After polarization, each transient acquisition starts following a magnetic $\pi$-pulse (corresponding to a $180$ degrees flip of the $^{13}$C spin along any direction). The external AC-magnetic field's phase varies between transient acquisitions (orange). As a result, the sidebands generally possess different phases in each transient spectrum. Averaging the transients yields a spectrum in which the sidebands are destructively averaged out (purple). Shifting each transient by a phase equal to the external field's accumulated phase restores the sidebands' phase coherence, yielding a spectrum with high signal-to-noise ratio sidebands (orange). For clarity, only one of the two Zeeman-split $J$-coupling peaks and its two sidebands are shown.  \textbf{Bottom}: result of the phase-shifting procedure for actual data. \textbf{(a) } Transients are averaged using $2001$ phase increments and stacked into a two-dimensional plot. \textbf{(b)  } Side view of \textbf{(a)}, sidebands are rescaled by a factor $10$ for  clarity. \textbf{(c) } Averaging with $\phi=0$ rad corresponds to averaging the transients without phase shift; sidebands are averaged out and carrier peaks appear with maximum amplitude. When the optimal phase (for $\omega/2\pi=0.73$ Hz, $\phi=2.93$ rad) is approached, sidebands appear. These spectra were acquired in an experiment during which the AC-field frequency and amplitude were set to $0.73$~Hz and $0.24$~nT. Transient acquisitions of $30$ s were repeated $850$ times with a time interval between each transient of $\tau=61$ s.}
		\label{fig:2dplot}	\label{fig:scheme}
	\end{figure*}
	
	\subsection{\textbf{Dark-matter signatures in zero- to ultralow-field NMR}}\label{ss:signatureZULF}

	The results presented in this work were obtained by applying techniques of ZULF NMR (see Materials and Methods Sec. \ref{s:experimentalsetup}, a review of ZULF NMR can be found in Ref. \cite{Blanchard2016d}).
	The sample -- $^{13}$C-formic acid, effectively a two-spin \mbox{${}^{1}$H--${}^{13}$C} system -- is pre-polarized  in a $1.8$~T permanent magnet and pneumatically shuttled to a magnetically-shielded environment for magnetization evolution and detection.

	The 
	spin Hamiltonian describing the system is
	\begin{multline}\label{2SpinHamiltonianwithALP}
	\mathcal{H} =  \hbar J\ts{CH} \bs{I} \cdot \bs{S}- \hbar \left(\gamma\ts{H} \bs{I} + \gamma\ts{C} \bs{S}\right) \cdot\bs{B} \\
	+\hbar \left(g\ts{app} \bs{I} - \frac{1}{3}g\ts{ann} \bs{S} \right) \cdot \bs{D}\ts{wind}(t)~,
	\end{multline}
	where the electron-mediated spin-spin coupling $J\ts{CH}/2\pi \approx 221~\mathrm{Hz}$ for formic acid and $\bs{{I}}$ and $\bs{{S}}$ are the nuclear-spin operators for $^{1}$H and $^{13}$C, respectively.
	Additionally, $\gamma\ts{H}$ and $\gamma\ts{C}$ are the gyromagnetic ratios of the  ${}^{1}$H and ${}^{13}$C spins,
	$\bs{B}$ is an applied magnetic field,
	$g\ts{app}$ is the ALP-proton coupling strength,
	and $g\ts{ann}$ is the ALP-neutron coupling strength.
	We assume $g\ts{app}=g\ts{ann}=g\ts{aNN}$ \cite{nEdmFlambaum}.
	
	In the absence of external fields ($\bs{B} = \bs{D}\ts{wind}(t) = 0$), the nuclear spin energy eigenstates are a singlet with total angular momentum $F=0$ and three degenerate triplet states with $F=1$, separated by $\hbar J\ts{CH}$.
	The observable in our experiment is the $y$ magnetization, leading to selection rules \mbox{$\Delta F = 0,\pm 1$} and \mbox{$\Delta m_F = \pm 1$,} as in Ref.~\cite{Ledbetter2011}.
	The zero-field spectrum thus consists of a single Lorentzian located at $J\ts{CH}/2\pi \approx 221~\rm{Hz}$, as shown in Fig.~\ref{fig:formicacid}(a).
	
	In the presence of a static magnetic field, $\bs{B}=B\ts{z}\bs{\hat{e}}\ts{z}$, applied along $z$, the $m_F=0$ states are unaffected, while the $m_F=\pm1$ triplet states' degeneracy is lifted.
	The corresponding spectrum exhibits two peaks at \mbox{$J\ts{CH}/2\pi \pm B\ts{z} (\gamma\ts{H}+\gamma\ts{C})/2\pi$}, as shown in Fig.~\ref{fig:formicacid}(b).
	
	So long as $|J\ts{CH}| \gg |\gamma\ts{H} B| \gg |g\ts{aNN} D\ts{wind}|$ the $m_F=0$ states are unaffected, and the $m_F=\pm1$ states are shifted by
	\begin{multline}
	\Delta E(m_F=\pm1)(t)=\mp\frac{\hbar}{2} B_z(\gamma\ts{H}+\gamma\ts{C})\\
	\pm \frac{\hbar}{2}\frac{2}{3}g\ts{aNN} D_z (t)~,
	\label{Eq:BALP-Splitting2}
	\end{multline}
	where $D\ts{z} (t)$ is the projection of $\bs{D}\ts{wind}(t)$ along the axis of the applied magnetic field.
	The time dependence of $\bs{D}\ts{wind}(t)$ leads to an oscillatory modulation of the \mbox{$m_F=\pm1$} energy levels, giving rise to sidebands around the $J$-coupling doublet as shown in Fig.~\ref{fig:formicacid}(c).
	The sidebands are separated from the carrier peaks by $\pm \omega_{\rm{DM}}/2\pi$ and have an amplitude proportional to the modulation index ($g_{\text{aNN}}/\omega_{\rm{DM}}$).
	
	Dark-matter fields with sufficiently strong coupling to nuclear spins can then be detected by searching for frequency-modulation-induced sidebands in the well-defined ZULF NMR spectrum of formic acid.

	\begin{figure*}	
		\centering
		\includegraphics{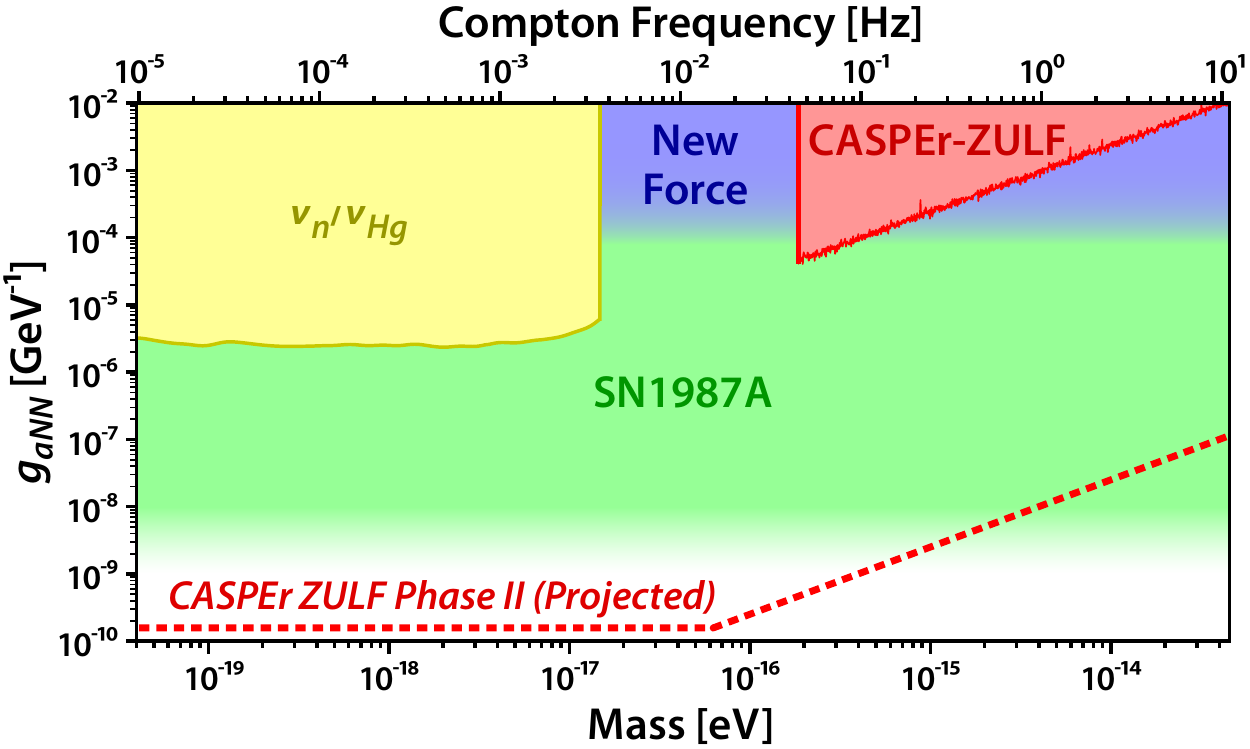}
		\caption{ALP wind-nucleon linear coupling parameter space. 
The CASPEr-ZULF region is excluded by this work ($90 \%$ confidence level) using a thermally polarized sample (data averaged over $850$ transient acquisitions of $30$ s each). The ``New Force'' region is excluded by searches for new spin-dependent forces \cite{spinDependentForces}. 
The SN1987A region represents existing limits from supernova SN1987A cooling \cite{Raffelt, Vysotsky}. The $\nu\ts{n}/\nu\ts{Hg}$ region  is excluded by measurements of the ratio of neutron and ${}^{199}$Hg Larmor precession frequencies \cite{nEdmFlambaum}. 
The dashed line corresponds to the sensitivity of a planned second phase of CASPEr-ZULF, with a projected $\sim 10^5$ factor increase in sensitivity  and the bandwidth extended towards lower frequencies by using a comagnetometer technique \cite{tengWu2019} and longer integration times.}
		\label{fig:ALPExclusion}	
	\end{figure*}	
	
	\subsection{\textbf{Coherent averaging}}
	
	The expected dark matter coherence time ($\sim14$~hours for a particle with $19$~Hz Compton frequency) is much longer than the nuclear spin coherence time in ${}^{13}$C-formic acid ($\sim10$~s).
	Taking advantage of this mismatch, we introduce the post-processing phase-cycling technique shown in Fig. \ref{fig:2dplot} (see Materials and Methods Sec. \ref{s:signalprocessing}), which consists of incrementally phase shifting the transient spectra and subsequently averaging them together.
	If the phase increment matches the phase accumulated by the oscillating field between each transient acquisition, the sidebands add constructively.
	This allows coherent averaging of the complex spectra, such that the signal-to-noise ratio scales as $N^{1/2}$, where $N$ is the number of transients. 
	Because the dark-matter Compton frequency is unknown, it is necessary to repeat this operation for a large number of different phase increments (at least many as the number of transient acquisitions).

	\subsection{\textbf{Calibration}}
	The energy shifts produced by $D\ts{z} (t)$ [Eq.~\eqref{Eq:BALP-Splitting2}] are equivalent to those produced by a real magnetic field with amplitude\\[-10pt]
	\begin{multline}
	B\ts{wind}(t)= \frac{2}{3} \frac{g\ts{aNN}}{\gamma\ts{H}+\gamma\ts{C}}\cdot D\ts{z} (t)\\	
	\propto \frac{g\ts{aNN}}{\gamma\ts{H}+\gamma\ts{C}}\sqrt{\rho_{\rm{DM}}} \sin{(\omega_{\rm{DM}}t)}  \bs{v} \cdot  \bs{\hat{e}}\ts{z}.
	\label{Eq:ALP-GeVtoB}
	\end{multline}
	Similar relationships for dark-photon and quadratic-wind couplings are provided in the Materials and Methods section.
	
	Based on Eq.~\eqref{Eq:ALP-GeVtoB}, the sensitivity of the experiment to dark matter was calibrated by applying a real oscillating magnetic field of known amplitude and frequency and measuring the amplitude of the resulting sidebands in the coherently averaged spectrum.
	Further details are provided in the Supplementary Materials.

	\subsection{\textbf{Search and analysis}}
	
	The dark matter search data were acquired and processed as described above, but without a calibration AC-magnetic field applied.
	
	For each Compton frequency, the appropriate phase increment is computed, which identifies the corresponding coherently averaged spectrum to be analyzed.
	The noise in the spectrum defines a detection threshold at the 90\% confidence level (further details in \ref{S:threshold} in the Supplementary Materials).
	When the signal amplitude at the given frequency is below the threshold, we set limits on the dark matter couplings to nuclear spins at levels determined by the calibration and effective-field conversion factors (see Materials and Methods Sec. \ref{s:effectiveFields}).
	If the signal is above the threshold, a more stringent analysis is performed by fitting the coherently averaged spectrum to a four-sideband model.
	When the fit rules out detection, the threshold level is again used to set limits.
	
	In case of an apparent detection, further repeat measurements would need to be performed to confirm that the signal is persistent and exhibits expected sidereal and annual variations.
	
	

	\subsection{\textbf{CASPEr-ZULF search results: constraints on bosonic dark matter}}\label{s:measurmentprocedure}
	
	The results of the CASPEr-ZULF search for axionlike particles are given in Fig. \ref{fig:ALPExclusion}.
	The frequencies presenting sharp losses in sensitivities at $0.21$, $1.69$, and $2.16$ Hz were the ones for which the nearest optimal phase increment was close to zero, thus presenting maximal-amplitude $J$-coupling peaks, raising the detection threshold (see discussion in Sec.~\ref{s:30} of the Supplementary Materials).
	The red-shaded area labeled ``CASPEr-ZULF'' corresponds to upper bounds on nuclear-spin couplings to dark matter consisting of ALPs at the $90\%$ confidence level. 
	This represents our current sensitivity limitation after $850$ $30$-second transient acquisitions using samples thermally polarized at $\sim1.8$~T. 
	The ``CASPEr-ZULF Phase II'' line corresponds to the projected sensitivity of a future iteration of this work that will use a more sensitive magnetometry scheme to measure a larger sample with enhanced (non-equilibrium) nuclear spin polarization.
	
\begin{figure}
		\centering
		\includegraphics{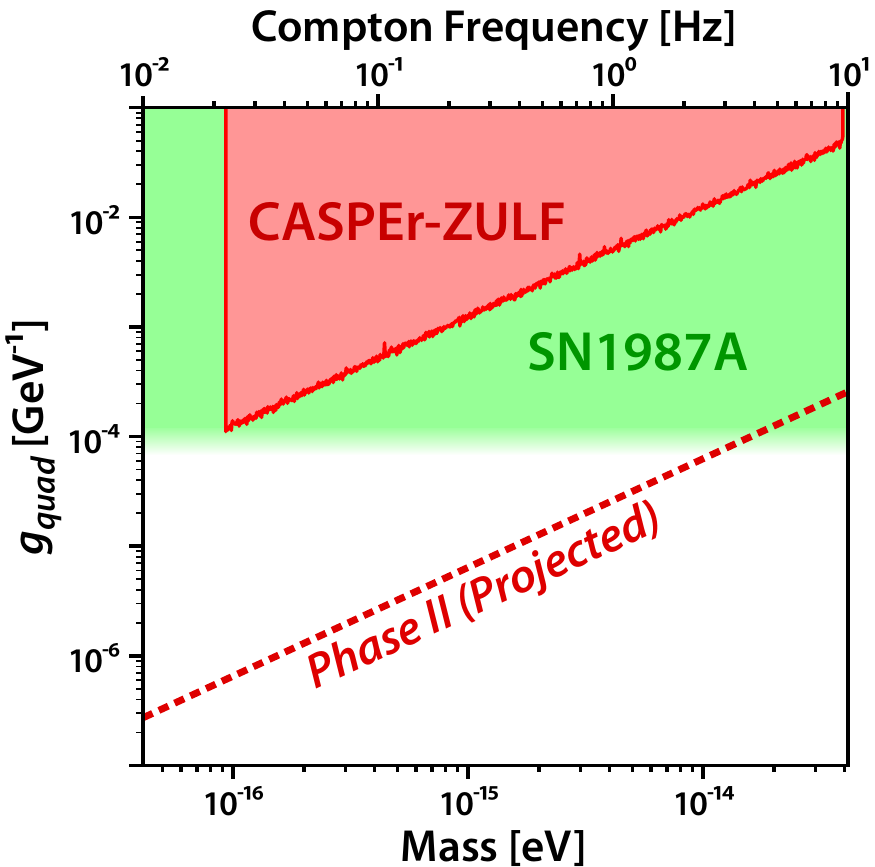}
		\caption{
			ALP wind-nucleon quadratic coupling parameter space.
			The CASPEr-ZULF region is excluded   by this work ($90 \%$ confidence level)  using a thermally polarized sample (data averaged over $850$ transient acquisitions of $30$ s each). 
			Other regions of this figure are defined in the caption of Fig. \ref{fig:ALPExclusion}.
		}
		\label{fig:QuadraticExclusionALP}
	\end{figure}

\begin{figure*}
		\centering
		\includegraphics{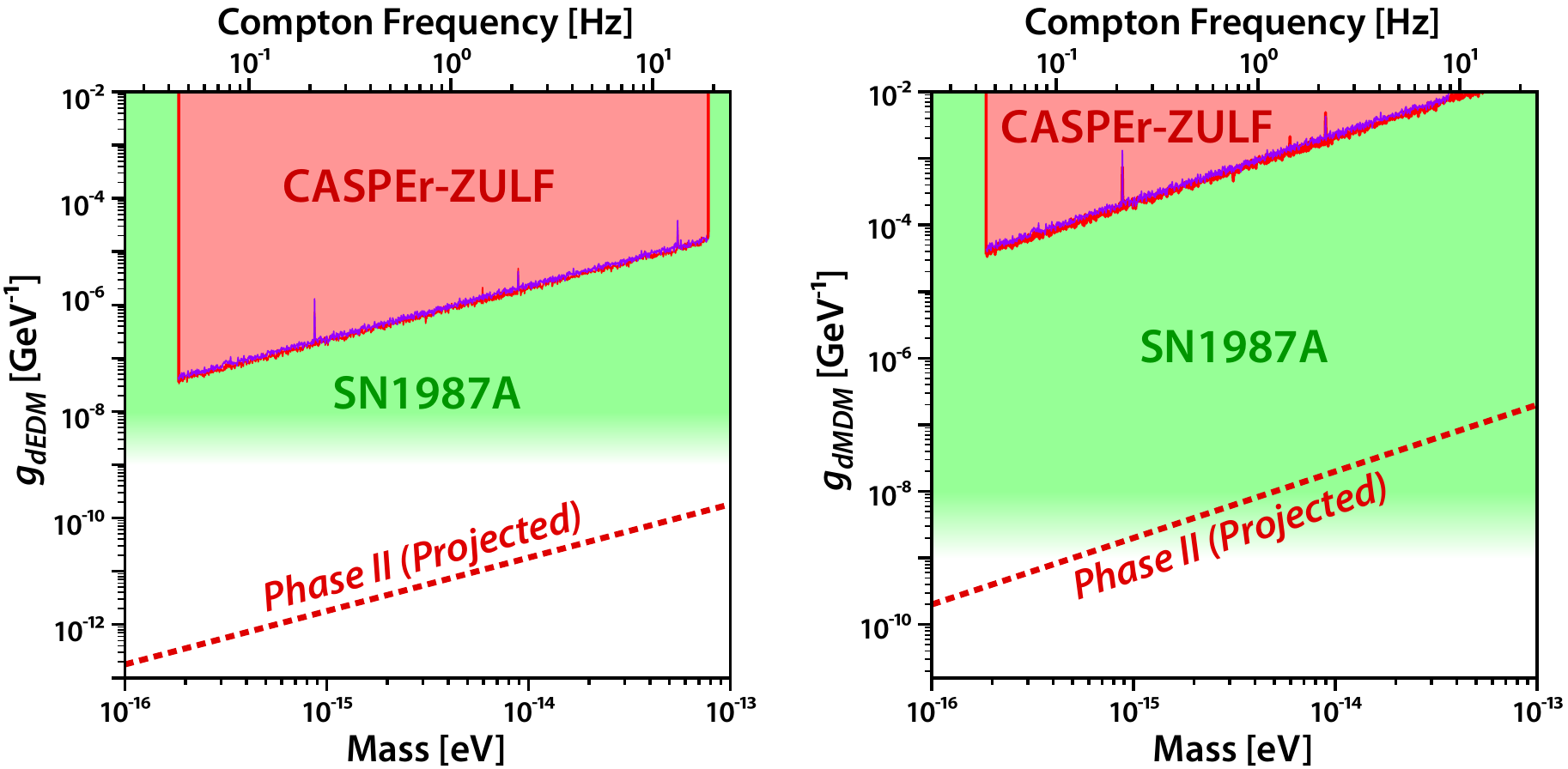}
		\caption{Left: dark photon-nucleon dEDM coupling parameter space.
			The SN1987A region represents existing limits for ALPs from supernova SN1987A cooling \cite{Raffelt, Vysotsky} adjusted to constrain dark photons as discussed in Ref. \cite{Gra17}. 
			Right: dark photon-nucleon dMDM coupling parameter space. 
			The CASPEr-ZULF regions are excluded by this work ($90 \%$ confidence level)  using a thermally polarized sample (data averaged over $850$ transient acquisitions of $30$ s each).
			The red and purple lines correspond to the case where the dark-photon field polarization is along the $\hat{\epsilon}_1$ and $\hat{\epsilon}_3$ axes of the non-rotating Celestial frame, respectively (see section \ref{s:dailyannualmodulation} of the Supplementary Materials).
			The dashed lines correspond to the sensitivity of a planned second phase of CASPEr-ZULF, with a projected $\sim 10^5$ factor increase in sensitivity.
		}
		\label{fig:PhotonExclusion}		
	\end{figure*}

	Figures \ref{fig:QuadraticExclusionALP} and  \ref{fig:PhotonExclusion} show the search results for the ALP quadratic interaction and dark photon interactions, respectively.
	No signal consistent with axion, ALP or dark-photon fields have been observed in the red-shaded areas.
	The two different limits given in Fig. \ref{fig:PhotonExclusion} were obtained using the same data set but analyzed by assuming two orthogonal initial polarizations of the dark-photon field (see \ref{s:dailyannualmodulation} in the Supplementary Materials). 
	
    In all cases, the search bandwidth was limited from below by the finite linewidth of the $J$-resonance peaks, preventing to resolve sidebands at frequencies lower than $45$~mHz.
	Due to the finite coherence time of the dark-matter fields (corresponding to $\sim10^6$ oscillations), the bandwidth's upper limit ($19$~Hz), is the highest frequency which can be coherently averaged after $14$~hours of integration time.
	The sensitivity fall off is due to the sidebands' amplitude scaling as the modulation index.
	Further details are given in \ref{S:calibration} and \ref{s:bandwith} of the Supplementary Materials.
	\noindent

	\section{\textbf{DISCUSSION}}\label{s:30}
	This work constitutes demonstration of a dark-matter search utilizing NMR techniques with a coupled heteronuclear spin system.
	The results provide new laboratory-based upper bounds for bosonic dark matter with masses ranging from \mbox{$1.86\times 10^{-16}$ to $7.85\times 10^{-14}$ eV}, complementing astrophysical bounds obtained from supernova SN1987A \cite{Raffelt,Vysotsky}.
	
	Our data analysis provides a method to perform coherent averaging of the bosonic-field-induced transient signals.
	This method should prove useful for other experiments seeking to measure external fields of unknown frequency using a detector with a comparatively short coherence time.
	Conveniently, this phase-cycling approach also suppresses the carrier-frequency signals, which would otherwise increase the detection threshold via spectral leakage.
	As this method is applied during post-processing, it does not require modification of the experiments provided that the data to be analyzed have been time stamped. 

	\subsection{\textbf{Phase II Sensitivity Improvements}}
	
	In order to search a greater region of the bosonic dark-matter parameter spaces, several sensitivity-enhancing improvements are planned for the next phase of the experiment. 
	In this work, nuclear spin polarization was achieved by allowing the sample to equilibrate in a $1.8$-T permanent magnet, which yields a ${}^{1}$H polarization $\lesssim 10^{-5}$.
	For the next phase of the experiment, a substantial sensitivity improvement can be obtained by using so-called ``hyperpolarization'' methods to achieve much higher, non-equilibrium nuclear-spin polarization. 
	Current efforts are focused on the implementation of non-hydrogenative parahydrogen-induced polarization (NH-PHIP)\footnote{Non-hydrogenative parahydrogen-induced polarization methods are often referred to with the acronym SABRE, for Signal Amplification by Reversible Exchange of parahydrogen.} \cite{SABRE}.
	Signal enhancement via NH-PHIP has been demonstrated at zero field \cite{Theis2012a} and after optimization is expected to increase nuclear spin polarization levels to at least 1\%.
	Because parahydrogen can be flowed continuously into the sample, a steady-state polarization enhancement can be achieved \cite{hovener2013hyperpolarized}, improving the experimental duty cycle.
	Combining continuous NH-PHIP with a feedback system to produce a self-oscillating nuclear spin resonator \cite{Suefke2017} could conceivably offer further improvement and simplify data analysis.
	
	Additional sensitivity enhancement will be provided by magnetometer improvements and use of a larger sample.
	In the experiments reported here, only about $50~\mathrm{\mu L}$ of the sample contributed to the signal, which was detected from below with an atomic magnetometer with a noise floor around $10~\mathrm{fT}/\sqrt{\mathrm{Hz}}$.
	With a larger ($\gtrsim 1$~mL) sample hyperpolarized via NH-PHIP detected via a gradiometric magnetometer array with optimized geometry and sensitivity below $1~\mathrm{fT}/\sqrt{\mathrm{Hz}}$, we anticipate an improvement by $\gtrsim10^5$ relative to the results presented here.
	
	Future experiments will be carried out with increased integration time.
	We recall that the bosonic dark-matter fields are coherent for a time $\tau\ts{DM}$  on the order of $10^6$ periods of oscillation.
	The phase-cycling procedure depicted in Fig.~\ref{fig:scheme} is valid for data sets with total time less than $\tau\ts{DM}$. 
	For integration times longer than the coherence time of the bosonic field \mbox{T$_{\text{tot}} > \tau\ts{DM}$}, the data sets have to be coherently averaged in sets of duration $\tau\ts{\rm{DM}}$ using the phase-cycling procedure. 
	This yields T$_{\text{tot}} / \tau\ts{DM} $ sets of coherently averaged data. 
	To  profit from longer integration time and further increase the SNR, these sets can be incoherently averaged by averaging their PSDs, yielding an overall SNR scaling as (T$_{\text{tot}} \tau\ts{DM}$)$^{1/4}$ (see Supplementary Materials in Ref. \cite{Bud14}).


	\subsection{\textbf{Complementary Searches}}
	
	In order to increase the bandwidth of the experiment (see Supplementary Materials \ref{s:bandwith}), we propose complementary measurement procedures. 
	As the amplitude of the sidebands scales as $1/\omega\ts{DM}$, the sensitivity of the experiment decreases for higher frequencies. 
	To probe frequencies ranging from $\sim 20$ to $500$ Hz (corresponding to bosonic masses of $\sim 8 \times 10^{-14}$ to $2 \times 10^{-12}$~eV), it was shown in Ref.  \cite{Gar17} that a resonant detection method will be more sensitive than the current frequency-modulation-induced sidebands measurement scheme. 
	Resonant AC fields can induce phase shifts in the $J$-coupling peaks \cite{Sjolander2016};
	cosmic fields can induce the same effect. 
	By gradually varying the magnitude of a leading magnetic field, one can tune the splitting of the $J$-coupling multiplets to match the dark-matter field frequency. 
	Such a resonance would manifest itself by shifting the phase of the $J$-coupling peak. 
	
	For frequencies below $\sim 45$ mHz (corresponding to bosonic masses $\lesssim 2 \times 10^{-16}$ eV), the sidebands are located inside of the $J$-coupling peaks and the experimental sensitivity drops rapidly. 
	This represents the lower limit of the bandwidth accessible by the frequency-modulation-induced sidebands measurement scheme presented in this work. 
	To probe down to arbitrarily low frequencies, another measurement scheme has been implemented based on a single-component liquid-state nuclear-spin comagnetometer \cite{Teng-Comag}.
	Further details and results of this scheme are presented elsewhere \cite{tengWu2019}.
	
	\section{\textbf{MATERIALS AND METHODS}}\label{s:40}
	
	\footnotesize
	
	\subsection{\textbf{Experimental parameters}}\label{s:experimentalsetup}
	The experimental setup used in this experiment is  the one described in Ref. \cite{billion1}. 
	Additional descriptions of similar ZULF NMR setups can be found in Refs \cite{ZULFreview}. 

	The NMR sample consists of $\sim 100$ $\mu$L of liquid $^{13}$C-formic acid (${}^{13}$CHOOH) obtained from ISOTEC Stable Isotopes (Millipore Sigma), degassed by several freeze-pump-thaw cycles under vacuum, and flame-sealed in a standard 5~mm glass NMR tube.
	
	The sample is thermally polarized for \mbox{$\sim 30$}~s in a $1.8$-T  permanent magnet, after which the NMR tube is pneumatically shuttled into the zero-field region.
	After the guiding magnetic field is turned off, a magnetic pulse (corresponding to a $\pi$ rotation of the $^{13}$C spin) is applied to initiate magnetization evolution.

	Following each transient acquisition, the sample is shuttled back into the polarizing magnet and the experiment is repeated.
	In order to increase the SNR, the transient signals are averaged using the phase-cycling technique described in Sec.~\ref{s:signalprocessing}.

	\subsection{\textbf{Bosonic dark-matter effective fields}}\label{s:effectiveFields}

	In the case of the ALP-wind linear coupling, the field acting on the $^{1}$H-$^{13}$C spins induces an energy shift equal to the one produced by a magnetic field with amplitude
\begin{multline}
B_{\mathrm{ALP,z}}^{\mathrm{eff}}(t)=-\frac{2}{3} \frac{g\ts{aNN}}{\gamma\ts{H}+\gamma\ts{C}}\sqrt{2 \hbar c \rho_{\rm{DM}}}\\
\times \sin(\omega\ts{DM} t - \bs{k}\cdot\bs{r} + \phi) \bs{v}\cdot \bs{\hat{e}}\ts{z},\label{eq:twospinwindmainbody}
\end{multline}
	where $\omega\ts{DM} \approx m\ts{DM}c^2/\hbar$ is the ALP Compton frequency, $\bs{k}\approx m\ts{DM} \bs{v}/\hbar$ is the wave-vector ($\bs{v}$ is the relative velocity), $m\ts{DM}$ is the rest mass of the ALP, $\phi$ is an unknown phase, and $\bs{\hat{e}}\ts{z}$ is the axis along which the leading DC-magnetic field is applied.
	
	It is theoretically possible that interaction of nuclear spins with $\bs{\nabla} a$ can be suppressed \cite{Oli08,Pos13}, in which case the dominant axion wind interaction, referred to as the quadratic wind coupling, is related to $\bs{\nabla} a^2$. 
	In the case of the ALP-wind quadratic coupling the equivalent magnetic field amplitude is:
\begin{multline}
 B_{\mathrm{quad,z}}^{\mathrm{eff}}(t)=-\frac{4}{3}  \hbar c^2\frac{g\ts{quad}^2}{\gamma\ts{H}+\gamma\ts{C}}\frac{\rho_{\rm{DM}}}{\omega_{\rm{DM}}}\\
\times \sin(2\omega\ts{DM} t - 2\bs{k}\cdot\bs{r} + \phi)\bs{v}\cdot \bs{\hat{e}}\ts{z}~. \label{quaddd}
\end{multline}
	where $g\ts{quad}$, having dimensions of inverse energy, parameterizes the ALP quadratic coupling strength to nuclear spins.
	
	There are two possible interactions of dark photons with nuclear spins that can be detected with CASPEr-ZULF:
	the coupling of the dark electric field to the dark electric dipole moment (dEDM) and 
	the coupling of the dark magnetic field to the dark magnetic dipole moment (dMDM). 
	The equivalent magnetic field amplitudes are:
\begin{multline}
 B_{\mathrm{dEDM}}^{\mathrm{eff}}(t) =\frac{2}{3} \frac{g\ts{dEDM}}{\gamma\ts{C}+\gamma\ts{H}} \sqrt{ 2 \hbar c^3 \rho\ts{DM} }\\
 \times \cos(\omega\ts{DM} t + \phi)\bs{\epsilon}\cdot \bs{\hat{e}}\ts{z}~,
\label{Eq:dark-photon-EDM-pseudo-field-EDM2mainbody}		
\end{multline}	
	and
\begin{multline}
 B_{\mathrm{dMDM}}^{\mathrm{eff}}(t) =\frac{2}{3} \frac{g\ts{dMDM}}{\gamma\ts{C}+\gamma\ts{H}} \frac{v}{c} \sqrt{ 2 \hbar c^3 \rho\ts{DM} }\\
  \times \cos(\omega\ts{DM} t + \phi)\bs{\epsilon}\cdot \bs{\hat{e}}\ts{z}~,\label{Eq:dark-photon-EDM-pseudo-field-MDM2mainbody}
\end{multline}
	with coupling constants $g\ts{dEDM}$ and $g\ts{dMDM}$ (having dimensions of inverse energy) and dark photon field polarization $\bs{\epsilon}$.
	
	The experimental sensitivity to real magnetic fields then directly translates to sensitivity to the coupling constants $g\ts{aNN}$, $g\ts{quad}$, $g\ts{dEDM}$, and $g\ts{dMDM}$. 
	Inverting equations \eqref{eq:twospinwindmainbody}--\eqref{Eq:dark-photon-EDM-pseudo-field-MDM2mainbody} yields the corresponding conversion factors from magnetic field to the dark-matter coupling constants:
\begin{align}
 \delta g\ts{aNN}(\omega)&\approx \left[1.3 \times10^{8} \phantom{a} \frac{\text{GeV}^{-1}}{\text{T}} \right] \phantom{a} \delta B(\omega), \label{conversion1}\\
\delta g\ts{quad}(\omega)&\approx \left[190 \phantom{a} \frac{\text{GeV}^{-1}}{\sqrt{\text{T}}\sqrt{\text{rad/s}}}\right] \phantom{a}\sqrt{\omega \cdot \delta B(\omega)} ,\label{conversion2}\\ 
\delta g\ts{dEDM}(\omega)&\approx \left[1.3 \times10^{5} \phantom{a} \frac{\text{GeV}^{-1}}{\text{T}}\right] \phantom{a} \delta B(\omega),  \label{conversion3} \\
\delta g\ts{dMDM}(\omega)&\approx \left[1.3 \times10^{8} \phantom{a} \frac{\text{GeV}^{-1}}{\text{T}}\right] \phantom{a} \delta B(\omega). \label{conversion4}
\end{align}
	Here we have used  $\gamma\ts{C}/2\pi= 10.70 $ MHz.T$^{-1}$ and $\gamma\ts{H}/2\pi=42.57$ MHz.T$^{-1}$, $v\approx10^{-3}c$ and $\rho\ts{DM}\approx0.4$ GeV/cm$^3$. The full derivation of these expressions is given in \ref{ss:22} of the Supplementary Materials.
	
	\subsection{\textbf{Signal processing}}\label{s:signalprocessing}
	For each transient acquisition, the sample is prepared in the same initial state, which determines the phase of the $J$-coupling peaks.
	When averaging the transients together, the $J$-coupling peaks' amplitude and phase remain constant, while the uncorrelated noise is averaged away, thus increasing the SNR as the square root of the total integration time $\text{T}_{\text{tot}}$, i.e. SNR$(\text{T}_{\text{tot}}) \propto \text{T}_{\text{tot}}^{1/2}$.
	
	However, the dark-matter-related information resides not in the $J$-coupling peaks, but in their sidebands.
	The external bosonic field oscillates at an unknown frequency and its phase at the beginning of each transient acquisition is unknown. 
	This phase directly translates into the phase of the sidebands in the transient spectra: while the phase of the $J$-coupling peaks is identical from one transient acquisition to another, the phase of the sidebands varies. 
	As a result, naively averaging the transient spectra averages the sidebands away, thus removing the dark-matter-related information from the resulting spectrum.
	
	Here we use a post-processing phase-cycling technique which enables coherent averaging of the spectra in the frequency domain, even for transient signals for which no obvious experimental phase-locking can be achieved due to the unknown frequency of the signal. 
	The method is similar to acquisition techniques in which an external clock is used to register the  times of the transient acquisitions and post-processing phase shifting of the transient signals is employed to recover the external field's phase \cite{TOPs, Qdyne}. 
	
	The method relies on the fact that the bosonic field's phase at the beginning of each transient acquisition is unknown but not random. 
	Indeed we recall that the bosonic fields remain phase coherent for $\sim 10^6$ oscillations, which for frequencies below $19$ Hz is longer than the total integration time ($14$ hours). 
	Thus, precise knowledge of the transient-signal acquisition starting times enables recovery of the phase of the bosonic field.
	
	A full description of this averaging method is given in the Supplementary Materials \ref{S:averaging}. 
	Each transient spectrum is incrementally phase shifted prior to averaging. 
	If the phase shift is equal to the phase accumulated by the bosonic field between two transient acquisitions, then the phase stability of the $J$-coupling peaks is shifted to their sidebands which can thus be coherently averaged.
	
	Considering that the frequency of the bosonic field is unknown, the correct phase shift is also unknown. 
	Thus, we repeat the operation for $2001$ different phase increments between $[-\pi,\pi]$, yielding $2001$ averaged spectra, one of which being averaged with the phase increment such that the sidebands are coherent.
	
	To demonstrate the viability of this method, a small magnetic field was applied with amplitude $0.24$ nT oscillating at $0.73$ Hz, to simulate a dark-matter field.
	Using this processing technique, the SNR of the sidebands scales as SNR$(\text{T}_{\text{tot}}) \propto \text{T}_{\text{tot}}^{1/2}$ (see Fig. \ref{fig:infoscaling} in the Supplementary Materials), as expected during a coherent averaging procedure. 
	This is a dramatic improvement over the alternative power-spectrum averaging (typically implemented for sets of incoherent spectra), which would yield a $ \text{T}_{\text{tot}}^{1/4}$ scaling.

	\subsection{\textbf{Calibration}}\label{s:signal processing}
	The experimental sensitivity is defined by the ability to observe dark-matter-induced sidebands above the magnetometer noise floor.
	In order to show that the sideband amplitude scales with the modulation index, $B/\omega$, and does not present unusual scalings due to experimental errors, two calibration experiments were performed.
	A first calibration was performed by varying the amplitude of the AC-field from $95$ to $310$~pT while holding the frequency constant at $\omega=2\pi\times0.73$~Hz.
	Then the amplitude was held at $160$~pT while varying the frequency, from $0.45$ to $1.7$~Hz.
	The results of this experiment are shown in Fig. \ref{fig:amplitudevsfrequency} in the Supplementary Materials.
	Similar experiments were performed to determine the minimum detectable frequency (see \ref{s:bandwith} in the Supplementary Materials).
	Based on this calibration, we can extrapolate the expected sidebands amplitude, $\text{A}_{s}(B,\omega)$, for any field of amplitude $B$ and frequency $\omega$:
	\begin{align}
	\text{A}_{s}(B,\omega) =5.53\times10^{-6}~ \text{rad.s}^{-1} \times \frac{B [\text{T}]}{\omega [\text{rad.s}^{-1}]}.\label{eq:threshold}
	\end{align}	
	Then the magnetometer noise level determines the smallest detectable driving field, which is then converted to dark-matter coupling bounds via eqs. \eqref{conversion1}-\eqref{conversion4}.

	\section{\textbf{REFERENCES}}\label{s:50}
	\bibliography{CASPEr-ZULF-20190317}
	
	{ \footnotesize \noindent
		\textbf{Acknowledgement}: The authors would like to thank Dionysis Antypas, Deniz Aybas, Andrei Derevianko, Martin Engler, Pavel Fadeev,  Matthew Lawson, Hector Masia Roig, and Joseph Smiga for useful discussions and comments. 	
		\textbf{Funding}: This project has received funding from the European Research Council (ERC) under the European Unions Horizon 2020 research and innovation programme (grant agreement No 695405). 
		We acknowledge the support of the Simons and Heising-Simons Foundations and the DFG Reinhart Koselleck project.
		PWG acknowledges support from DOE Grant DE-SC0012012, NSF Grant PHY-1720397, DOE HEP QuantISED award \#100495, and the Gordon and Betty Moore Foundation Grant GBMF794.
		DFJK acknowledges the support of the U.S. National Science Foundation under grant number PHY-1707875. 
		YVS was supported by the Humboldt Research Fellowship.
		\textbf{Authors contribution}: 
		JWB, DB, DFJK and AOS conceived the project; 
		JWB oversaw and managed the progress of the project;
		JWB and TW designed, constructed and calibrated the experimental apparatus and software controls. JWB, DB and AG conceived the data acquisition scheme; 
		AG acquired and analysed the data with contribution of JWB, GC, NLF  and TW;
		AG wrote the manuscript with the contribution of JWB, DB, GC, NLF, DFJK,  AW and TW;
		PWG, DFJK, SR and YVS conceived the theory related to axion and hidden photon coupling to nuclear spins. 
		All authors have read and contributed to the final form of the manuscript.	
		\textbf{Competing interests}: the authors declare that they have no competing interests. 
		\textbf{Data and materials availability}: all data needed to evaluate the conclusions in the paper are present in the paper and/or the Supplementary Materials. Additional data related to this paper may be requested from the authors.}

	\cleardoublepage
	\newcommand{\beginsupplement}{%
		\setcounter{table}{0}
		\renewcommand{\thetable}{S\arabic{table}}%
		\setcounter{figure}{0}
		\renewcommand{\thefigure}{S\arabic{figure}}%
		\setcounter{section}{0}
		\renewcommand{\thesection}{S\arabic{section}}
	}
	
	\begin{centering}
		{ Supplementary Materials for \\ 
			\small \textbf{Constraints on bosonic dark matter from ultralow-field nuclear magnetic resonance}}
		
		{\vspace{0.5cm}\small Antoine~Garcon, John~W.~Blanchard, Gary~Centers,  Nataniel~L.~Figueroa, Peter~W.~Graham, Derek~F.~Jackson~Kimball, Surjeet~Rajendran, Alexander~O.~Sushkov, Yevgeny~V.~Stadnik, Arne~Wickenbrock, Teng~Wu, and~Dmitry~Budker}\vspace{0.5cm}
	\end{centering}
	\beginsupplement
	
	\small
	
	\textbf{The PDF file includes:}\vspace{0.5cm}\\
	•~Fig.~\ref{fig:amplitudevsfrequency}. Calibration data, signal amplitude scaling with respect to external fields frequency and amplitude.\\
	•~Fig.~\ref{fig:fitcdf}.~Noise analysis data, signal threshold definition\\
	•~Fig.~\ref{fig:infoscaling}.~Signal amplitude, signal detection threshold and signal-to-noise ratio scaling with respect to integration time. \\	
	•~Fig.~\ref{fig:bandwidthEnd}.~Signal amplitude scaling with respect to external fields frequency. Low frequency analysis to define bandwidth lower bound. \\
	•~References~\cite{billion1, Tayler2016, Blanchard2016d, romalisSERF, Bud2006, Gra17, Gar17, Gra13,nedm13,Gra15review, Oli08,Pos13, Scargle1982, Bandyopadhyay2012}.

	\section{Experimental setup}\label{S:setup}
	The apparatus used in this experiment is identical to that described in Ref. \cite{billion1}. 
	Additional descriptions of similar ZULF NMR setups can be found in Ref. \cite{Tayler2016}.
	
	The zero-field region is achieved using a four-layer $\mu$-metal and ferrite shield (MS-1F shield produced by Twinleaf LLC, $10^6$ magnetic shielding factor). 
	In this region, the residual magnetic field is $\lesssim 10$ pT. 
	Three orthogonal pairs of Helmholtz coils  which receive current from an amplified source (AE Techron 7224-P) are wound on the holder. 
	This allows for the application of AC and DC magnetic fields along three orthogonal directions. 
	This set of coils is used to apply acquisition-starting pulses.  A separate set of coils mounted on the innermost shield is used to apply the shimming, leading DC-, calibration AC-, and benchmark AC-fields.
	
	Prior to every transient acquisition, the sample is thermally polarized for $\sim30$  s in a $1.8$ T permanent  magnet,  after  which  the  NMR  tube  is  pneumatically shuttled into the zero-field region.  
	During the shuttling, the sample goes through a magnetic field of a guiding solenoid wrapped around a fiberglass tube,  ensuring adiabatic transport into the shielded region \cite{Blanchard2016d}. 
	
	After the shuttling to the zero-field region, the sample stops $\sim 1$ mm 
	above the spin-exchange-relaxation-free magnetometer's (SERF \cite{romalisSERF,Bud2006}) rubidium-vapor cell (cell produced by Twinleaf with $500$ torr of nitrogen buffer gas). 
	
	In order for the magnetometer to operate in SERF regime, the rubidium cell is maintained at $180$ C$^{\circ}$ by means of a resistive heater.  
	The $795$ nm circularly-polarized pump beam (from a Toptica Photonics diode laser system, locked to the D$1$ Rubidium line), propagates along the y-axis.
	
	The linearly polarized probe beam (from another Toptica Photonics diode laser system), propagating along the x-axis, is blue-detuned $10$ GHz from the center of the rubidium D$2$ line to probe the rubidium atoms' polarization. 
	The polarimeter's analog output signal is digitized with a 24 bit acquisition card (NI 9239, National Instrument) at a $5$ kHz sampling rate. 
	Synchronization of the experiments and control of shuttling, magnetic pulses, and data acquisition is accomplished with a LabVIEW program. 
	
	
	\section{\textbf{Measurement schemes}}\label{Measurementschemes}
	
	\subsection{\textbf{Zero-field regime}}\label{s:ZULFNMR}
	$^{13}$C-formic acid possesses an electron-mediated spin-spin coupling between the $^{13}$C and its neighbouring  $^{1}$H, referred to as $J$-coupling. In isotropic liquids, rotational motion of the molecules averages the $J$-coupling Hamiltonian to its isotropic form:
	\begin{align}
	\mathcal{H}\ts{J}=\hbar  J\ts{CH}  \bs{I}\cdot\bs{S}~, \label{eq:ZULFhamiltonian}
	\end{align}
	where $J\ts{CH}/2\pi $ $\approx$ $221$ Hz for formic acid, $\bs{I}$ and $\bs{S}$ are the nuclear spin-1/2 operators for $^{1}$H and $^{13}$C, respectively. 
	
	Prior to the acquisition, the sample is shuttled to the magnetically shielded region where the residual magnetic field is on the order of $10$ pT. In such a low magnetic field, the Zeeman interaction is on the order of $0.1$ mHz and is negligible compared to the $J$-coupling. For isotropic fluids in zero-field, other interactions typically relevant in NMR spectroscopy, such as the short-range dipole-dipole couplings, are averaged-out by molecular tumbling and are also negligible. As a result, the sample's evolution in zero-field is dominated by the $J$-coupling Hamiltonian given by Eq. (\ref{eq:ZULFhamiltonian}). 
	
	Following shuttling, the guiding field is turned off suddenly and a magnetic field pulse with area $\gamma_C B_x t_{\mathrm{pulse}}=\pi$ is applied to the sample. 
	As a result, the $^{13}$C and $^1$H nuclear spins are in a superposition of the triplet $\ket{F\text{=}1; m_F\text{=} \pm 1, 0}$ and singlet $\ket{F\text{=}0; m_{F}\text{=} 0}$ states which evolves under the $J$-coupling Hamiltonian. 
	The unperturbed energy levels are given by:
	\begin{equation}
	E(F)/\hbar= \frac{1}{2} J\ts{CH}[F(F+1) - I(I+1) - S(S+1)],
	\end{equation}
	where $I=S=1/2$ for $^{13}$C and $^{1}$H and $F=0,1$. Then the singlet and triplet states energy levels are respectively:
	\begin{align}
	&E(F=0, m_F=0)/ \hbar= - \frac{3}{4}J\ts{CH}~,\nonumber \\ 
	&E(F=1, m_F=0,\pm 1)/ \hbar=\frac{1}{4}J\ts{CH}~.
	\end{align}

	The selection rules for transition between the singlet and triplet states, \mbox{$\Delta F = 0, \pm 1$ and $\Delta m_F = \pm 1$}, arise because the observable is the magnetization along the y-axis. The corresponding spectrum exhibits a single peak centred at \mbox{$\Big[E(F=1,\pm 1)-E(F=0)\Big]/ \hbar= J \approx 2\pi\cdot 221$} Hz (see Fig. \ref{fig:formicacid}.a of the main text).

	\subsection{\textbf{Ultralow-field regime}}\label{s:DC}	
	The experimental setup allows application of a DC-magnetic field, $\bs{B}\ts{z}$, along the z-axis via the Helmholtz coils surrounding the sample. In such conditions, the Hamiltonian under which the formic acid molecules evolve becomes:
	\begin{align}
	\mathcal{H}&=	\mathcal{H}\ts{J} + \mathcal{H}\ts{Z}\\
	&=	\mathcal{H}\ts{J} -  \hbar ( \gamma\ts{C} {S}\ts{z} +  \gamma\ts{H} {I}\ts{z}) B\ts{z}~,
	\end{align}
	where $\gamma\ts{C}/2\pi= 10.70 $ MHz.T$^{-1}$ and $\gamma\ts{H}/2\pi=42.57$ MHz.T$^{-1}$ are the gyromagnetic ratios of the $^{13}$C and $^1 $H nuclear spins, respectively. A field of about  $50$ nT yields a Zeeman interaction, $H\ts{Z}$, on the order of $ \gamma_{C}{B}\ts{z}/2\pi \approx \gamma_{H}{B}\ts{z}/2\pi \approx 0.1  $ Hz, which remains much weaker than the $J$-coupling and can thus be treated as a perturbation.  To first order in ${B}\ts{z}$, the eigenstates are those of the unperturbed Hamiltonian, $\mathcal{H}\ts{J}$, and perturbed energies can be read from the diagonal matrix elements of the Zeeman perturbation:
	\begin{equation}
	\Delta E(F,m_F)/\hbar=	-\bra{F,m_F}(\gamma\ts{C} {S}\ts{z} + \gamma\ts{H} I\ts{z}) B\ts{z}\ket{F,m_F},
	\end{equation}
	yielding the perturbed energy levels:
	\begin{align}
	&E(F=0, m_F=0)/\hbar= -  \frac{3}{4}J\ts{CH}~,\nonumber \\ 
	&E(F=1, m_F=0,\pm 1)/\hbar= \frac{J\ts{CH}}{4} -   m_F \frac{\gamma\ts{C} +  \gamma\ts{H}}{2} B\ts{z}~. \label{DCperturbationEnergy}	
	\end{align}	
	Thus, to first order, the effect of such a field is to break the degeneracy of the triplet state due to the now non-negligible Zeeman splitting. Recalling the selection rules $\Delta F = 0, \pm 1$ and $\Delta m_F = \pm 1$, the magnetometer measures oscillations at two different frequencies between the $\ket{\text{F=}0; \text{m}\ts{F}\text{=} 0}$ and $\ket{\text{F=}1; \text{m}\ts{F}\text{=} \pm 1}$ states. As a result, the single $J$-coupling line is split into a doublet and the spectrum exhibits two lines located at \mbox{$J/2\pi  \pm B\ts{z}\frac{\gamma_\text{C}  + \gamma_\text{H} }{4\pi}$} (see Fig. \ref{fig:formicacid}.b of the publication main text)

	\subsection{\textbf{Frequency-modulation regime }}\label{ss:22}
	In this regime, we apply an oscillating magnetic field $\bs{B}\ts{AC} =  B\ts{AC} \cos(\omega t)\cdot\bs{\hat{e}}\ts{z}$ in addition to the DC-field along z-axis. Under these conditions, the formic acid molecules evolve under the following Hamiltonian:
	\begin{align}
	\mathcal{H}&=	\mathcal{H}\ts{J} + \mathcal{H}\ts{Z} + \mathcal{H}\ts{AC}\\
	&=\mathcal{H}\ts{J} + \mathcal{H}\ts{Z}-\hbar ( \gamma\ts{C} S\ts{z}+   \gamma\ts{H} I\ts{z}) B\ts{AC}\cos(\omega  t)~.
	\end{align}
	
	In the case where the modulation index, $\frac{ B\ts{AC}(\gamma_\text{C}  + \gamma_\text{H})}{2\omega }  $, is much smaller than one, the resulting spectrum is composed of the $J$-coupling doublet peaks, located at \mbox{$J\ts{CH}/2\pi  \pm B\ts{z}\frac{\gamma_\text{C}  + \gamma_\text{H} }{4\pi}$}  each exhibiting a set of sidebands located at  frequencies \mbox{$J\ts{CH}/2\pi  \pm B\ts{z}\frac{\gamma_\text{C}  + \gamma_\text{H} }{4\pi} \pm \omega /2\pi$}.  Such a spectrum is shown in Fig.~\ref{fig:formicacid}.c of the publication main text, alongside the corresponding energy levels. The sidebands' amplitude $\text{A}\ts{s}$, is proportional to the modulation index \cite{Gar17}. This behaviour was experimentally confirmed by varying the amplitude and frequency of a calibration AC-field while a  DC-field was continuously applied to the sample. This procedure yields the calibration data, necessary to interpret results from the search for dark-matter bosonic fields.
	
	\section{\textbf{Dark matter effective fields}}\label{ss:33}
	In the following, we explicitly state the Hamiltonians describing the interaction of ALPs and dark photons with a single nuclear spin-1/2 having gyromagnetic ratio $\gamma\ts{N}$. By analogy to the Zeeman interaction, we then derive the expressions of the corresponding pseudo-magnetic fields acting on a two-spin system such as formic acid. In the following, all expressions are given in SI units.
	
	\subsection{\textbf{ALP Wind linear coupling to nuclear spins}}\label{ss:lin}
	For a single nuclear spin, axions and ALPs generate a pseudo-magnetic field through what is known as the “axion wind interaction”, described by the nonrelativistic Hamiltonian \cite{Gra13,nedm13,Gra15review},	
	\begin{align}
	\mathcal{H}\ts{wind} &= \sqrt{\hbar^{3} c^{3}} g\ts{aNN} \bs{\nabla} a(\bs{r},t) \cdot \bs{{I}}\ts{N}\\
	&= -\hbar g\ts{aNN} \bs{D}\ts{wind} \cdot \bs{I}\ts{N}~,
	\end{align}
	where:
	\begin{align}
	\bs{D}\ts{wind}(t) = -  \sqrt{\hbar c^{3}} \bs{\nabla} a(\bs{r},t)~,
	\end{align}
	is the ALP effective field, acting on the nuclear spin, and $ \bs{I}\ts{N}$ is the spin operator in units of $\hbar$. 
	
	The ALP field, $a(\bs{r},t)$, can be written as:
	\begin{align}
	a(\bs{r},t)= a_0 \cos(\omega\ts{DM} t - \bs{k}\cdot\bs{r} + \phi)~,\label{axionfield}
	\end{align}
	where $\omega\ts{DM} \approx m\ts{DM}c^2/\hbar$ is the ALP Compton frequency, $\bs{k}\approx m\ts{DM} \bs{v}/\hbar$ is its wave-vector with $\bs{v}$, the average velocity of the particles in the laboratory frame, $m\ts{DM}$ is the rest mass of the ALP particle and $\phi$ is an unknown phase.
	As the value of $\phi$ has no incidence on the measurement we set its value to zero for the rest of this discussion.
	Differentiating Eq. \eqref{axionfield} yields:	
	\begin{align}
	\bs{\nabla} a(\bs{r},t)& =  \frac{m\ts{DM} a_0}{\hbar} \sin(\omega\ts{DM} t - \bs{k}\cdot\bs{r}) \bs{v}~,
	\end{align}
	
	Recalling that $a_0$ is related to the dark-matter density through
	\begin{equation}
	\rho\ts{DM} = \frac{1}{2 }\frac{c^2}{\hbar^2} m\ts{DM} ^2 a_0^2~,\label{eq:DMdensityagain}
	\end{equation}
	yields:
	\begin{equation}
	\nonumber \bs{D}\ts{wind}(t) =-  \sqrt{2  \hbar c \rho\ts{DM} }
	\times \sin(\omega\ts{DM} t - \bs{k}\cdot\bs{r}) \bs{v}.
	\end{equation}
	
	We now assume that the ALP field is acting on the nucleons of the coupled $^{13}$C and $^1$H nuclear spins while a weak DC magnetic field is applied to the sample. The axion wind Hamiltonian now becomes
	\begin{align}\label{Eq: 2SpinHamiltonianwithALP2}
	\mathcal{H}\ts{wind} = - \hbar  \bs{D}\ts{wind}\cdot \left(g\ts{app} \bs{I} - \frac{1}{3}g\ts{ann} \bs{S} \right)  ~,
	\end{align}
	where, $g\ts{app}$ is the ALP-proton coupling strength, $g\ts{ann}$ is the ALP-neutron coupling strength. To first order, the $m_F=0$ states are unaffected, and the $m_F=\pm 1$ states are shifted by
	\begin{equation}
	\Delta E(m_F=\pm1)(t)=\pm \frac{\hbar}{2}\frac{2}{3}g\ts{aNN} D\ts{z} (t)~,
	\label{Eq:BALP-Splitting}
	\end{equation}
	where we have assumed that $g\ts{app}=g\ts{ann}=g\ts{aNN}$, and $D\ts{z} (t)$ is the projection of $\bs{D}\ts{wind}(t)$ along the axis of the applied magnetic field. Thus the perturbation induced by $D\ts{z} (t)$ takes a similar form to that of a magnetic field $B\ts{z}(t)$ [see Eq. \eqref{DCperturbationEnergy} of the publication main text]:
	\begin{align}
	B_z(t)&= \frac{2}{3}\frac{g\ts{aNN} }{\gamma\ts{H}+\gamma\ts{C}} D\ts{z} (t)\\
	\nonumber &=-\frac{2}{3} \frac{g\ts{aNN}}{\gamma\ts{H}+\gamma\ts{C}}\sqrt{2  \hbar c \rho\ts{DM} } \sin(\omega\ts{DM} t - \bs{k}\cdot\bs{r}) \bs{v}\cdot \bs{\hat{e}}\ts{z}~.\label{eq:twospinwind}
	\end{align}
	
	\subsection{\textbf{ALP Wind quadratic coupling to nuclear spins}}\label{ss:quad}	
	It is theoretically possible that interaction of nuclear spins with $\bs{\nabla} a$ can be suppressed \cite{Oli08,Pos13}, in which case the dominant axion wind interaction, referred to as the quadratic wind coupling, is related to $\bs{\nabla} a^2$:  
	\begin{align}
	\mathcal{H}\ts{quad}&= \hbar^2 c^2 (g\ts{quad})^2 \bs{\nabla}a^2(\bs{r},t)\cdot \bs{I}_\text{N}\\
	&= -\hbar (g\ts{quad})^2 \bs{D}\ts{quad} \cdot \bs{I}_\text{N}~,
	\label{Eq:quadratic-wind-hamiltonian}
	\end{align}
	where $g\ts{quad}$, having dimensions of inverse energy, parameterizes the ALP quadratic coupling strength to nuclear spins and where we have introduced the effective field due to the quadratic axion coupling:
	\begin{align}
	\bs{D}\ts{quad}&=-   \hbar c^2 \bs{\nabla}a^2(\bs{r},t)~.
	\end{align}
	Assuming the form of $a(r,t)$ given by Eq. \eqref{axionfield}:
	\begin{align}
	\bs{\nabla}a^2(\bs{r},t)&=a_0^2 \bs{k}\sin(2\omega\ts{DM} t - 2\bs{k}\cdot\bs{r})~,
	\end{align}
    we obtain for the effective field:
    \begin{equation}
     \bs{D}\ts{quad}(t)=-2   \hbar c^2\frac{\rho_{\rm{DM}}}{\omega_{\rm{DM}}}\sin(2\omega\ts{DM} t - 2\bs{k}\cdot\bs{r})\bs{v}~.
	\end{equation}	 
	As for the linear coupling, the effect of $\bs{D}\ts{quad}$ acting on a single spin is equivalent to that of a  magnetic field, whose amplitude in this case is given by: 
	\begin{align}
	B\ts{quad,z}(t)&=\frac{g\ts{quad}^2}{\gamma\ts{N}} {D}_\text{quad,z}(t)\\
	\nonumber&= -2 \hbar c^2 \frac{g\ts{quad}^2}{\gamma_N} \frac{ \rho_{\rm{DM}}}{\omega_{\rm{DM}}} \times\sin(2\omega\ts{DM} t - 2\bs{k}\cdot\bs{r})\bs{v}\cdot \bs{\hat{e}}\ts{z}~. \label{Eq:quadratic-wind-pseudo-field}
	\end{align}
	Specifically for formic acid, assuming equal coupling of the axion field to the proton and the neutron, we have
	\begin{multline}
	 B\ts{quad,z}(t)=-\frac{4}{3} 
	\hbar c^2 \frac{g\ts{quad}^2}{\gamma\ts{H}+\gamma\ts{C}}\frac{\rho_{\rm{DM}}}{\omega_{\rm{DM}}}\\
	\times \sin(2\omega\ts{DM} t - 2\bs{k}\cdot\bs{r})\bs{v}\cdot \bs{\hat{e}}\ts{z}~.
	\end{multline}
	We note that for the quadratic coupling case, the modulation frequency is twice the Compton frequency of the ALPs. Therefore, an experiment sensitive to a certain range of axion masses for the linear coupling is, at the same time, probing quadratic couplings of axions of half the mass.
	
	\subsection{\textbf{Dark-photon couplings to nuclear spins}}\label{ss:dp}
There are two possible interactions of dark photons with nuclear spins that can be detected with CASPEr-ZULF:
	the coupling of the dark electric field to the dark electric dipole moment (dEDM) and 
	the coupling of the dark magnetic field to the dark magnetic dipole moment (dMDM). 
	
	Assuming the dark photons make up the totality of the dark matter energy density, $\rho_{\rm{DM}}$, the energy stored in the dark electric field, $\bs{\bs{E}}\ts{dEDM}(t)$, is set equal to $\rho_{\rm{DM}}$, in a form analogous to that of the usual electromagnetism: 
	\begin{align}
	\rho_{\rm{DM}}=\frac{\epsilon_0}{2}|\bs{\bs{E}}\ts{dEDM}|^2~,
	\end{align}
	where $\epsilon_0$ is permittivity of free space. The dark magnetic field is given by:
	\begin{align}
	|\bs{\bs{B}}\ts{dMDM}|\approx \frac{v}{c}|\bs{\bs{E}}\ts{dEDM}|~.
	\end{align}
	The Hamiltonians describing the dark EDM and dark MDM are respectively given by:
	\begin{align}
	\mathcal{H}\ts{dEDM} &=  g\ts{dEDM} \sqrt{\frac{\hbar^3c^3}{\pi}} \bs{\bs{E}}\ts{dEDM}(t) \cdot \bs{I}\ts{N}~,\label{Eq:Dark-photon-EDM-Hamiltonian}	
	\end{align}		
	and: 
	\begin{align}
	\mathcal{H}\ts{dMDM} &= g\ts{dMDM} \sqrt{\frac{\hbar^3c^3}{\pi}} \bs{\bs{B}}\ts{dMDM}(t) \cdot \bs{I}\ts{N}~,\label{Eq:Dark-photon-EDMMDM-Hamiltonian}
	\end{align}	
	where $g\ts{dEDM}$ and $g\ts{dMDM}$ (of dimension of inverse energy) parametrize the strength of the dark photon couplings to the EDM and MDM, respectively. 	
	
	Similarly to axions and ALPs, the above Hamiltonians are expressed by analogy to the Zeeman interaction by introducing the following magnetic fields:
	\begin{equation}
	 B\ts{dEDM}(t) = \frac{g\ts{dEDM}}{\gamma\ts{N}} \sqrt{ 2 \hbar c^3 \rho\ts{DM} }  \cos(\omega\ts{DM} t)\bs{\hat{\epsilon}}\cdot \bs{\hat{e}}\ts{z}~,
	\label{Eq:dark-photon-EDM-pseudo-field-EDM}		
	\end{equation}	
	and
	\begin{equation}
	 B\ts{dMDM}(t) = \frac{g\ts{dMDM}}{\gamma\ts{N}} \frac{v}{c} \sqrt{ 2 \hbar c^3 \rho\ts{DM} }\cos(\omega\ts{DM} t)\bs{\hat{\epsilon}}\cdot \bs{\hat{e}}\ts{z}~,\label{Eq:dark-photon-EDM-pseudo-field-MDM}
	\end{equation}
	where $\bs{\hat{\epsilon}}$ is the bosonic field's polarization vector.
	Assuming equal coupling to the proton and neutron, yields for a two-spin ($^1$H-$^{13}$C) system:
	\begin{equation}
	 B\ts{dEDM}(t) = \frac{2}{3} \frac{g\ts{dEDM}}{\gamma\ts{C}+\gamma\ts{H}} \sqrt{ 2 \hbar c^3 \rho\ts{DM}} \cos(\omega\ts{DM} t)\bs{\hat{\epsilon}}\cdot \bs{\hat{e}}\ts{z}~,
	\label{Eq:dark-photon-EDM-pseudo-field-EDM2}		
	\end{equation}	
	and
	\begin{equation}
 B\ts{dMDM}(t) = \frac{2}{3} \frac{g\ts{dMDM}}{\gamma\ts{C}+\gamma\ts{H}} \frac{v}{c} \sqrt{ 2 \hbar c^3 \rho\ts{DM} } \cos(\omega\ts{DM} t)\bs{\hat{\epsilon}}\cdot \bs{\hat{e}}\ts{z}~,\label{Eq:dark-photon-EDM-pseudo-field-MDM2}
	\end{equation}
	Note that for ordinary photons, the coupling to the EDM  is strongly suppressed since the electromagnetic interaction respects the CP symmetry. However, the dark photon need not respect the CP symmetry, in which case there would be no associated suppression \cite{Gra17}.

	\section{Signal processing: post-processing phase-cycling}\label{S:averaging}
	The averaging method is illustrated in Fig. \ref{fig:scheme} in the publication main text. Subsequent transient acquisitions are separated by a time interval  $\tau$. The $n^{\text{th}}$ transient acquisition yields a transient spectrum denoted FFT$_{n}$.  The operation is repeated $N$ times, yielding a set of $N$ transient spectra: \mbox{$\Big\{\text{FFT}_{n} $ where ${n}=1,2...$  ${N}\Big\}$.} Once every $\text{FFT}_{n}$ has been acquired, a phase shift, $\phi$, incremented by the current acquisition number $n$, is  applied to each $\text{FFT}_{n}$:
	\begin{align}
	\text{FFT}_{n} ^{\phi} := \text{FFT}_{n} \times \exp(-i n  \phi)~.
	\end{align}
	The $N$ transient phase-shifted $\text{FFT}_{n} ^{\phi}$ are then averaged:
	\begin{align}
	\langle\text{FFT} ^{\phi}\rangle  := \frac{1}{{N}} \times \sum_{n=1} ^{N} \text{FFT}_{n} ^{\phi}~.
	\end{align}
If the phase shift matches the accumulated phase of the dark-matter field between each transient acquisition,
	\begin{align}
	\phi := \int_{0}^{\tau}\omega\ts{DM}\text{dt} \phantom{qqwe}(\textbf{modulo }2 \pi),
	\label{optimalPhase}
	\end{align}
	where $\omega\ts{DM}$ is the oscillation frequency of the dark-matter field, the sidebands will be averaged coherently.
	Generally, the ``carrier'' $J$-coupling peaks will be averaged to zero, with the exception of cases where $\omega_{DM} = 2\pi/\tau$. 
	Because the sidebands are coherently averaged, their SNR scales as SNR$(\text{T}_{\text{tot}}) \propto \text{T}_{\text{tot}}^{1/2}$ while the $J$-coupling peaks are averaged away along with the uncorrelated noise.
	
	The data shown in Fig. \ref{fig:2dplot} of the main text were acquired in an experiment during which the AC-field frequency was set to $0.73$~Hz. 
	The time interval between each measurement was set to  $\tau=61$~s (including $30$~s of acquisition and $31$~s for polarization, shuttling and pulsing, combined). The optimal phase shift is computed using Eq. \eqref{optimalPhase}, yielding $\phi \approx 2.93$ rad.	Note that as the frequency of the bosonic fields is unkown, this optimal phase shift cannot be computed during the actual dark-matter-search run. However, this optimal phase shift must necessarily lie within the [$-\pi, \pi$] interval. In practice we numerically try $2001$ phase increments within [$-\pi, \pi$] : \mbox{$\Big\{\phi_k=k\pi / 1000 $~where~$k=-1000,-999,...1000\Big\}$.}. This method produces a set of $2001$ phase-shifted averaged FFTs: $\langle\text{FFT}^{\phi_k}\rangle$. 

	Figure \ref{fig:2dplot} illustrates the result of the phase-shifting processing method for an experiment where the transient acquisition is repeated $850$ times.
	The $\langle\text{FFT}^{\phi_k}\rangle$ are ordered by phase increment $\phi_k$, and stacked in a two-dimensional plot. This plot is then examined and searched for sidebands. The phase increment of $\phi=0$ rad corresponds to averaging without applying any phase shift. In this case, the averaging is optimized for the $J$-resonances and sidebands cannot be seen (i.e., they are destructively averaged). However, a specific phase increment of $\phi = 2.93 $ rad (highlighted in orange), yields maximum sideband amplitude while the $J$-coupling peaks are averaged out. This phase value (and its negative counterpart) corresponds to the optimal value for which the benchmark-field coherence is restored and yields maximum SNR, matching the computed value using Eq. \eqref{optimalPhase}.

	\section{\textbf{Sideband amplitude determination}}\label{sss:423}
	A benchmark sideband measurement was performed using a test field of amplitude \mbox{$B_0=0.24$ nT}, oscillating at the frequency \mbox{$\omega_0/ 2\pi=0.73$ Hz.} The transient acquisition data, consisting of  \mbox{$30$ s} time-series sampled at \mbox{$5$ kHz}, are Fast-Fourier Transformed. The transient FFTs are multiplied by a calibration function expressing the magnetometer sensitivity within the \mbox{$0$ to $500$ Hz} bandwidth. 
	The calibration function is obtained by measuring the magnetometer response to an applied frequency-varying magnetic field of fixed amplitude. 
	
	Following the calibration, the baseline of the FFTs is fit and subtracted. 
	The transient FFTs are then averaged using the phase-shifting scheme described in the previous section.
	
	The next operation consists in extracting the sideband amplitudes from the averaged spectrum. To this end, six complex Lorentzians are fit to the averaged spectrum (two for the $J$-peaks, each possessing two sidebands):
	\begin{align}
	L(\nu) &:=  \frac{\alpha}{ \Gamma -i (\nu - \nu_\text{c})}\exp{(i\phi)}~,
	\end{align}
	where $\alpha$, $\Gamma$, $\nu_\text{c}$ and $\phi$ are the amplitudes, linewidths, center frequencies and phases of the Lorentzians, respectively, and are real, freely fitted parameters. 
	The benchmark sideband amplitude, centered at $\nu_\text{c}$ is given by $\text{A}_{s_0}=|L(\nu_\text{c})|$.
	
	\section{Calibration: signal scaling versus bosonic-field amplitude and frequency}\label{S:calibration}
	In order to experimentally verify that the sideband amplitude scales with the modulation index, $\text{A}_s(\text{B},\omega) \propto \text{B}/\omega$, two calibration experiments were performed. 
	
\begin{figure*}
		\centering
		\includegraphics[width=0.98\linewidth]{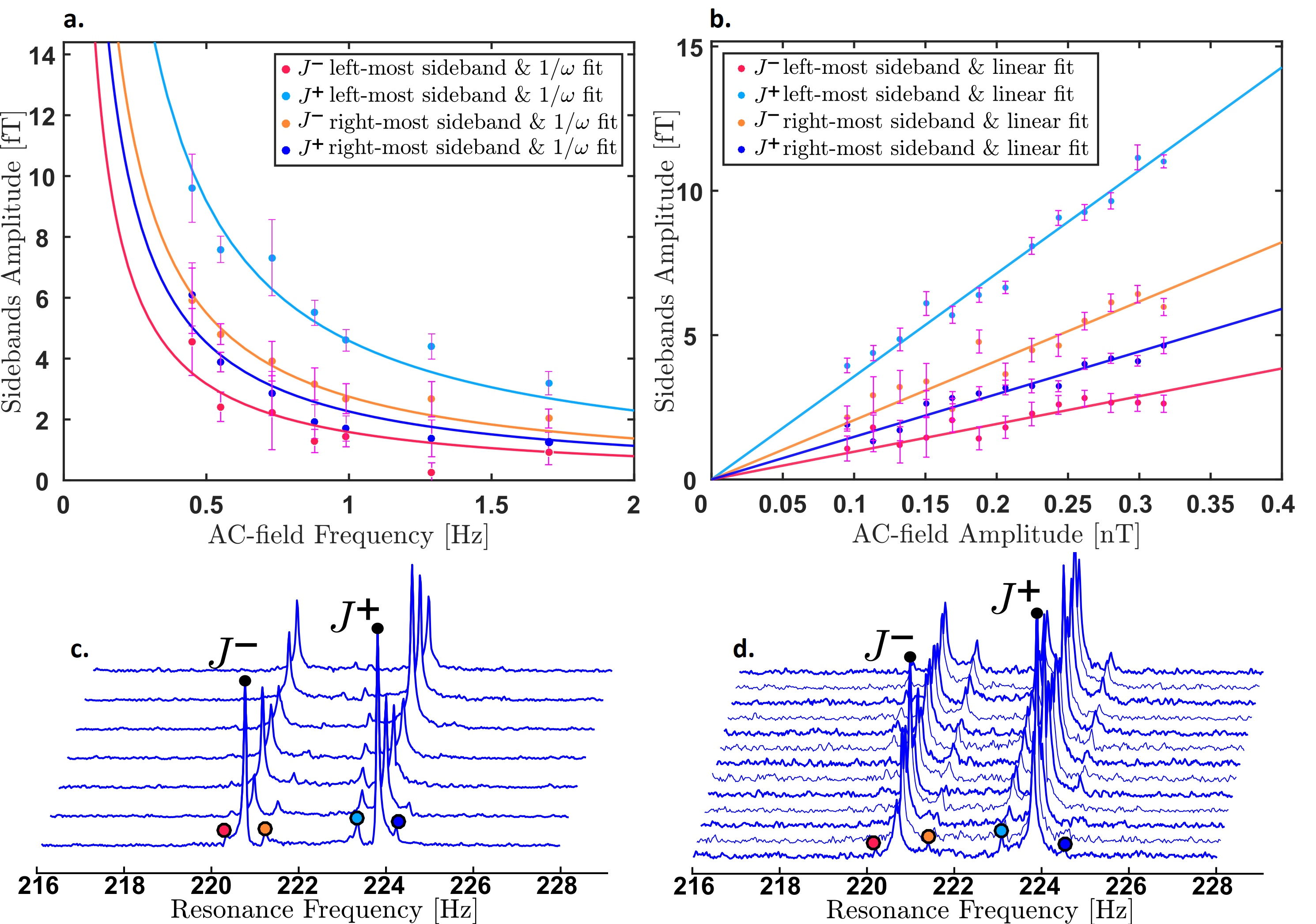}
		\caption{ Calibration data.
			\textbf{(a) \& (b) }: Sideband amplitude scaling with external field amplitude and frequency.
			\textbf{(a) \& (c) }: Calibration-field set to $0.16$ nT and frequency, $\omega/2\pi$, varied from $0.45$ to $1.7$ Hz. The sideband amplitude follows a $1/\omega$ dependence on the calibration-field frequency $\omega$. 
			\textbf{(b) \& (d) }: Calibration-field frequency set to $0.73$ Hz and amplitude varied from \mbox{$0.095$ to $0.31$ nT}. The sideband amplitude follows a linear dependence on the calibration-field amplitude. 
			\textbf{(a) \& (b) }: Each point is the amplitude of the sideband lorentzian fit. Data are averaged over $70$ scans of $30$ s using the optimal phase increment corresponding to the probed frequency. Error bars show the $1$-$\sigma$ confidence interval obtained from the fitting procedure.}
		\label{fig:amplitudevsfrequency}
		\label{fig:amplitudevsamplitude}
	\end{figure*}

	A first calibration was performed by varying the amplitude of the AC-field with constant frequency. The results of this experiment are shown in Fig. \ref{fig:amplitudevsfrequency} and show the linear dependence of the sideband amplitude on the amplitude of the AC-field. Each data point corresponds to the measured sideband amplitude (as defined is the previous section) for AC-calibration fields of amplitude varying from \mbox{$0.095$ to $0.31$ nT} at fixed frequency $0.73$ Hz. 
	
	The second calibration was performed by varying the frequency of the AC-calibration field from $0.45$ to $1.7$ Hz with constant amplitude of $0.16$ nT. The results of this calibration (Fig. \ref{fig:amplitudevsfrequency}) show the $1/\omega$ dependence of the sideband amplitude with the frequency of the AC-field.
	
	Both calibration experiments were carried out with $2100$ seconds total integration time  for each data point (corresponding to  $70$ transient acquisitions of  $30$ seconds).
	\section{Detection threshold determination}\label{S:threshold}
\begin{figure*}
		\centering
		\includegraphics[width=0.90\linewidth]{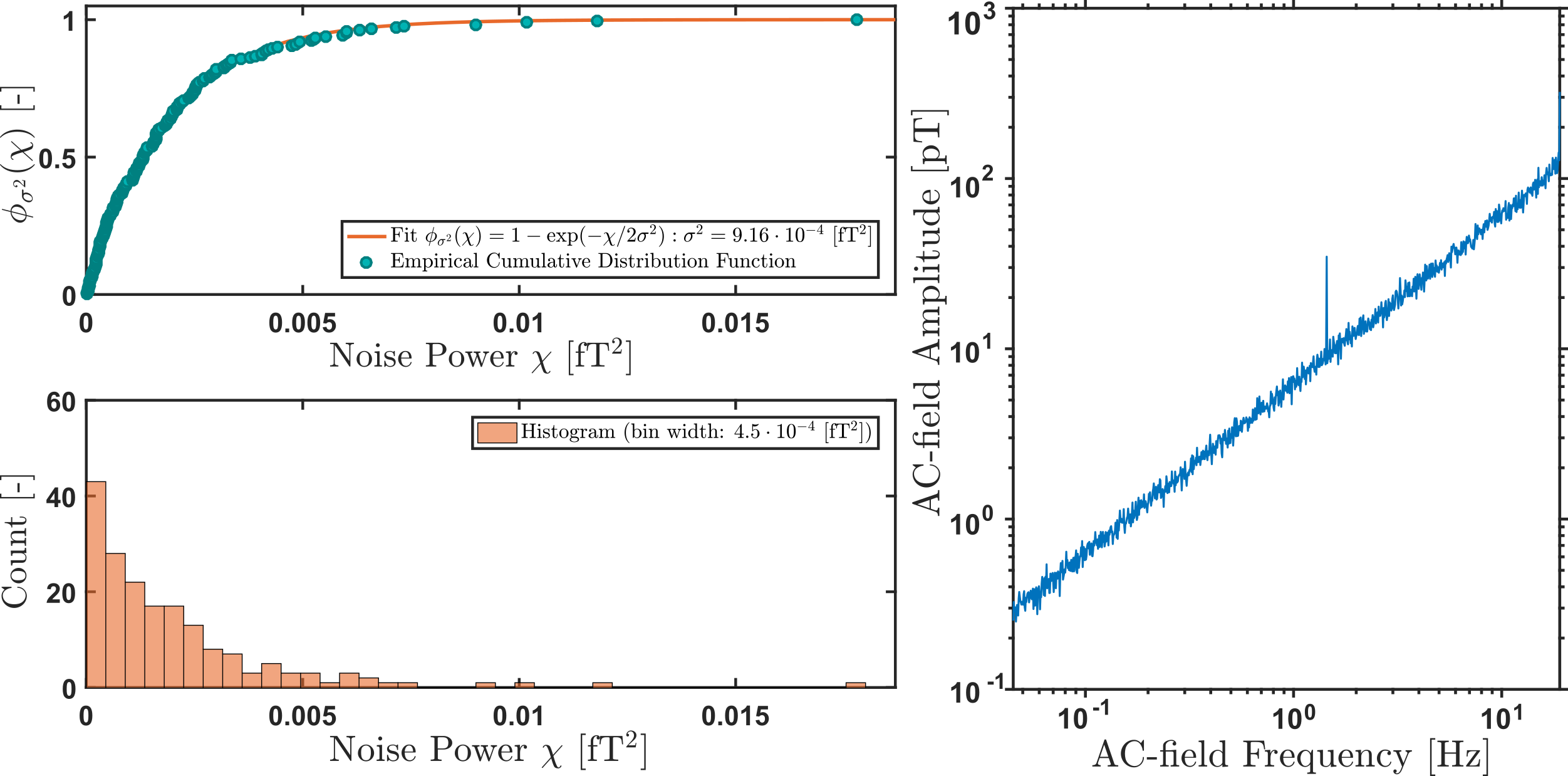}
		\caption{\textbf{(a)} : Example of empirical white-noise power-spectral-density cumulative distribution function (CDF) and fit to the exponential distribution \mbox{$\phi_{\sigma^2}\{\chi\} = 1 - e^{-\chi/2\sigma^2}$.} The fit yields a white-noise variance $\sigma^2 \approx 9.16 \times 10^{-4}$ [fT$^2$]. \textbf{(b)} : White-noise power histogram. \textbf{(a), (b)} : Non-corrected for misalignment (see \ref{s:dailyannualmodulation}), noisy-data analysed around $1.5$ Hz in a $5$ Hz frequency bin using the optimal phase increment corresponding to the $1.5$ Hz frequency. \textbf{(c)} : Amplitude of AC magnetic field yielding a signal above the detection threshold for frequencies within the bandwidth $0.045$ to $19$ Hz. Based on the calibration data (see section \ref{S:calibration}) and power threshold evaluation. Total integration time of $25500$ s  (corresponding to $850$ transient acquisitions each $30$ seconds long, with a duty cycle of $50$\% imposed by the polarization step between each acquisition).}
		\label{fig:fitcdf}
	\end{figure*}

	The detection threshold, $\text{A}\ts{th}$, is defined as the required signal amplitude such that we can claim detection with a $90\%$ confidence in the case: $\text{A}_{s}(\text{B},\omega) >$ $\text{A}\ts{th}$, where $\text{A}_{s}(\text{B},\omega) $ is the measured signal as described in Eq. \eqref{eq:threshold}. To define the detection threshold, we use the standard $p$-value test described in Ref. \cite{Scargle1982}. To this aim we consider a complex spectrum containing noise only:
	\begin{align} 
	\text{FFT}(\omega) = \chi_1(\omega) + i \chi_2(\omega)~,
	\end{align}
	where $\chi_1(\omega)$ and  $\chi_2(\omega)$ are the real and imaginary part of the FFT, respectively, and $\omega/2\pi$ is the probed frequency. The signal detection threshold can be defined by studying the power spectral density (PSD) of this noisy spectrum:
	\begin{align}
	\text{PSD}(\omega) \propto \chi_1(\omega)^2 +\chi_2(\omega)^2~.
	\end{align}
	To this aim we use the result that if the Fourier coefficients  $\chi_1(\omega)$ and  $\chi_2(\omega)$ are normally distributed with zero mean, variance $\sigma^2$ and are frequency independent (consistent with white noise), then the PSD cumulative distribution function (CDF) follows an exponential distribution \cite{Scargle1982}. Hence, at any frequency $\omega/2\pi$, the probability that the measured power P is smaller than $z_0$ is:
	\begin{align}
	\text{Pr}\{P(\omega)<z_0\} = 1 - e^{-z_0/2\sigma^2}, \phantom{aa}\forall\omega~.\label{CDF}
	\end{align}
	In practice, the value of $\sigma^2$ is obtained by fitting the evaluating the PSD CDF of a noise-only spectrum (in the present case via the MATLAB Kaplan-Meier estimator). We then fit the PSD CDF to the exponential distribution given in Eq. \eqref{CDF}. An illustration of this procedure is shown in Fig. \ref{fig:fitcdf}.
	
	We now define a signal power $z_0$ such that, if at the probed frequency $\omega/2\pi$, the power of a spectrum containing both the signal and noise, P$(\omega)$, is greater that $z_0$, a detection can be claimed with a false-alarm probability $p_0$ (a false alarm corresponds to the case where the measured power P$(\omega)$ is due to noise fluctuations). This yields:
	\begin{align}
	p_0	:&= \text{Pr}\{\text{P}(\omega)>z_0\}\\
	&= 1- \text{Pr}\{\text{P}(\omega)<z_0\}\\
	&=  e^{-z_0/2\sigma^2}~.
	\end{align}
	Solving the above equation for $z_0$ yields:
	\begin{align}
	z_0 = -2\sigma^2 \ln(p_0)~.
	\end{align}
	Setting now the false-alarm probability to $p_0 = 0.10$ (corresponding to a confidence level of $90\%$) we find the signal power threshold is terms of the noise fluctuation:
	\begin{align}
	z_0 \approx 4.6 \sigma^2~.
	\end{align}
	As our analysis is done on the spectrum amplitude, defined as the square-root of the PSD, we define the detection threshold as:
	\begin{align}
	\text{A}\ts{th}& = \sqrt{z_0}\\
	& \approx 2.14 \sigma~.
	\end{align}
	Any signal with amplitude $\text{A}\ts{s}(B,\omega)>\text{A}\ts{th}$ is inspected for detection. The case where no such signal is observed, yields the upper-bound exclusion line:
	\begin{align}
	\text{A}_{s}(B,\omega) = \text{A}\ts{th}  ~ .
	\end{align}	
	Using Eq. \eqref{eq:threshold} this limit can be expressed in term of magnetic field versus frequency:
	\begin{align}
	B\ts{th}(\omega) = \frac{ \text{A}\ts{th}}{ \text{A}_{s_0} } \frac{B_0}{\omega_0} \omega ~.\label{BfieldvsFreq}
	\end{align}

	Such analysis is done in a frequency-dependent manner by computing the phase increment corresponding to the studied frequency. The averaged spectrum obtained using this phase increment is then studied as explained above to define the signal-detection threshold at $\omega$. The results of this analysis are given in Fig.~\ref{fig:fitcdf}.c, where Eq.~\eqref{BfieldvsFreq} is evaluated for every frequency within the bandwidth.
	
	The exclusion plots shown in Figs. \ref{fig:ALPExclusion}, \ref{fig:QuadraticExclusionALP} and \ref{fig:PhotonExclusion} in the publication main text, are obtained by converting $\omega$ to bosonic masses in eV and $B\ts{th}(\omega)$ to coupling constants in GeV$^{-1}$ using Eqs. \eqref{conversion1}, \eqref{conversion2}, \eqref{conversion3} and \eqref{conversion4} in the publication main text.
	
	The exclusion lines show the 2.14$\sigma$ confidence level of the experiment ($90 \%$ CL). Such frequency-dependent analysis yields higher detection threshold for frequencies for which the phase increment is close to $0$ rad. Indeed, the spectra obtained from the $0$ radian phase increment contains maximum amplitude $J$-couplings, which then raise the detection threshold to a higher value due to spectral leakage. This effect manifests itself in the exclusion lines by sporadic and sharp increases of the detection threshold.

	We note that in the literature, similar analyses are done by applying a statistical penalty for searching multiple frequencies within a bandwidth of interest to account for the look-elsewhere effect  (corresponding to raising Eq. \eqref{CDF} to the power of the number of frequencies probed). Our analysis does not require such a penalty. Indeed, the characteristic signature of the bosonic fields, a set of four sidebands around the $J$-coupling frequencies, allows one to differentiate false alarms from true detection events.
	
	\section{Coherent averaging: signal scaling with integration time}\label{S:coherent}
\begin{figure}
		\centering
		\includegraphics[width=.98\linewidth]{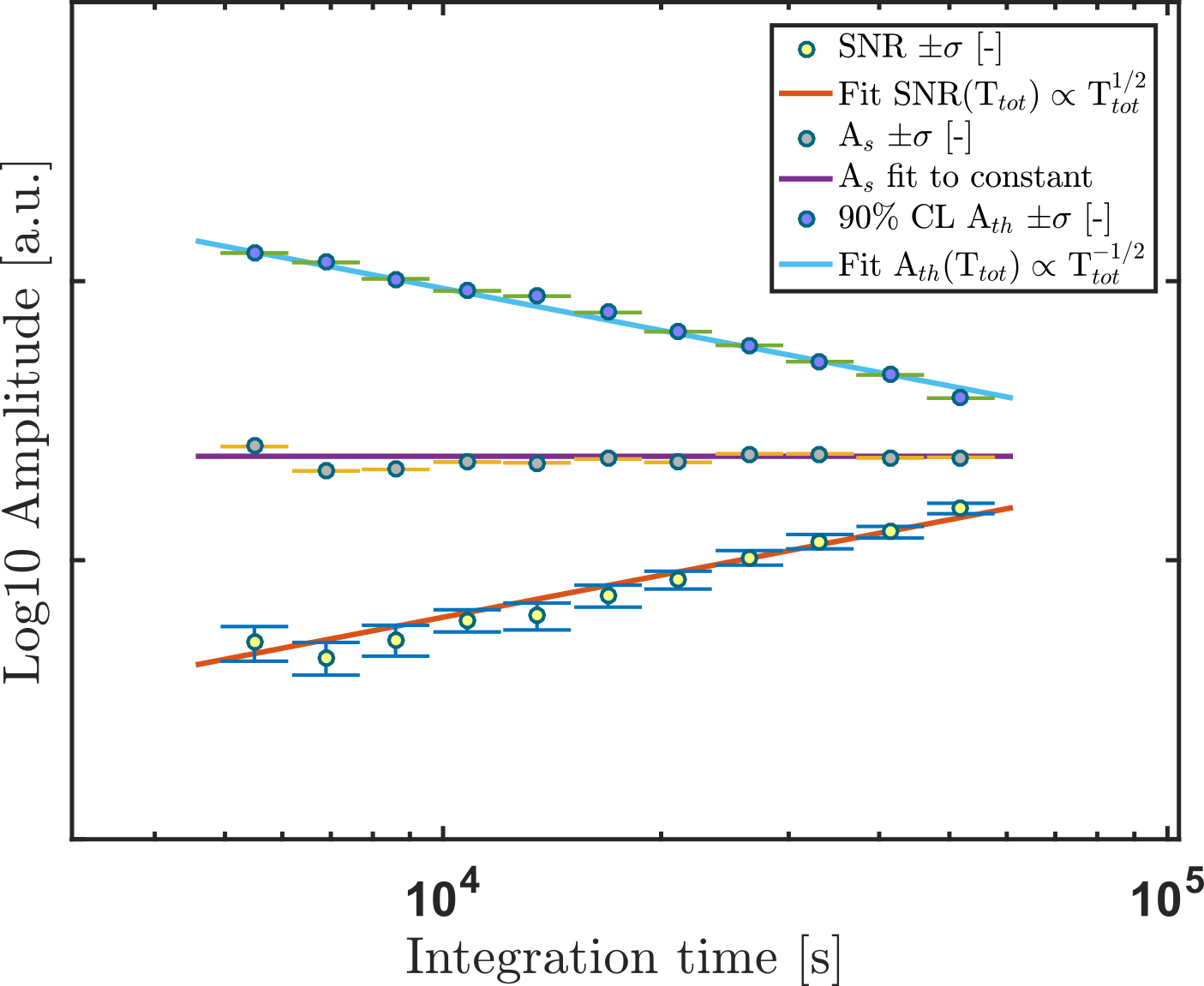}
		\caption{Sidebands amplitude $\text{A}\ts{s}$, detection threshold $\text{A}\ts{th}$ at $90\%$ confidence level ($90\%$ CL), and signal-to-noise ratio $\text{A}\ts{s}/\text{A}\ts{th}$ (SNR) versus integration time T$\ts{tot}$, using the phase shifting processing method. $\text{A}\ts{s}$ remains constant, $\text{A}\ts{th}$ scales as T$\ts{tot}^{-1/2}$  during the averaging, consistent with  coherent averaging. Thus the SNR scales as T$\ts{tot}^{1/2}$. During this experiment, the AC-field is maintained at a frequency $\omega\ts{AC}/2\pi=0.73$ Hz and amplitude $B\ts{AC}=0.24$ nT. Noisy data analysed around $0.73$ Hz in a $5$ Hz frequency bin. Data is averaged over $850$ scans of $30$ seconds using the optimal phase increment corresponding to the $0.73$ Hz frequency. Error bars show the $1$-$\sigma$ confidence interval obtained from the fitting procedure.}
		\label{fig:infoscaling}	
	\end{figure}

	Figure \ref{fig:infoscaling} shows the sidebands amplitude $\text{A}\ts{s}$, detection threshold  $\text{A}\ts{th}$, and signal-to-noise ratio, SNR:= $\text{A}\ts{s}/\text{A}\ts{th}$, scaling with the total integration time using the phase-shifting procedure described in \ref{S:averaging}. 
	
	During this experiment, the AC-field is maintained at a frequency \mbox{$\omega\ts{AC}/2\pi=0.73$ Hz} and amplitude ${B}\ts{AC}=0.24$ nT. 
	The experiment is repeated with increased number of transient acquisitions each of length $30$ s. All values are extracted using the averaged spectrum obtained from the optimal phase increment corresponding to  $\omega\ts{AC}$.
	
	Figure \ref{fig:infoscaling} shows that the sideband amplitude remains constant while the signal detection-thresholds decreases as  $\text{A}\ts{th}($T$\ts{tot})\propto$ T$\ts{tot}^{-1/2}$ as expected while performing a coherent averaging. The SNR scales accordingly as SNR $\propto$ T$\ts{tot}^{1/2}$.

	\section{\textbf{Bandwidth: accessible bosonic mass range}}\label{s:bandwith}
\begin{figure}
		\centering
		\includegraphics[width=0.98\linewidth]{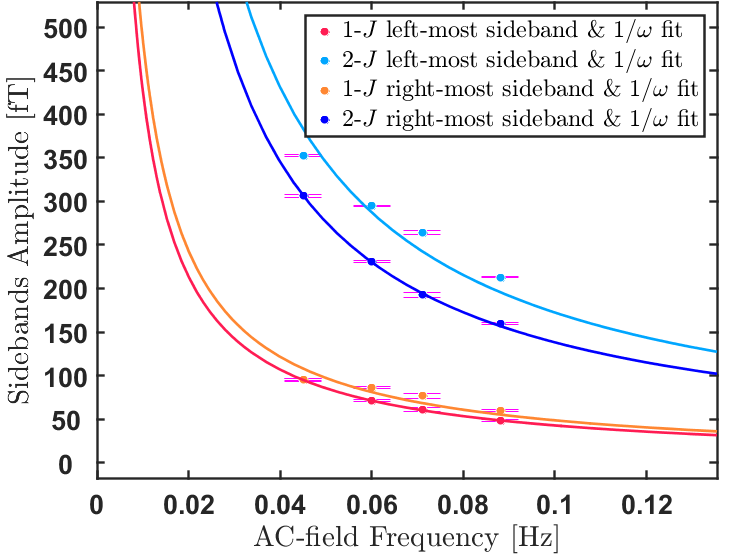}
		\caption{Sideband amplitude scaling with calibration-field frequency. Calibration-field set to $0.8$ nT and frequency, $\omega/2\pi$, varied from $0.045$ to $0.088$ Hz. The sideband amplitude follows a $1/\omega$ dependence with the calibration-field frequency $\omega/2\pi$. Each point is the amplitude of the sideband lorentzian fit. Data are averaged over $70$ scans of $30$ s using the optimal phase increment corresponding to the probed frequency. Error bars show the $1$-$\sigma$ confidence interval obtained from the fitting procedure.}
		\label{fig:bandwidthEnd}
	\end{figure}

	The bandwidth of the experiment is limited from below by the linewidth of the $J$-coupling peaks. For bosonic fields with frequencies lower than $45$ mHz  (corresponding to bosons with mass $\lesssim 1.86 \times 10^{-16}$ eV), the corresponding sidebands are located inside the $J$-resonance and cannot be resolved. As a result, we limit the lower end of the search bandwith to $45$ mHz, below which calibration could not be achieved (see Fig. \ref{fig:bandwidthEnd}). In principle, the bandwidth extends to frequencies up to $\sim 500$ Hz, corresponding to the maximum detectable frequency of the magnetometer (see Materials and Methods Sec. \ref{s:experimentalsetup}). However, the data are acquired during $\sim 14$ hours (corresponding to $850$ transient acquisitions each $30$ seconds long, with a duty cycle of $50$\% imposed by the polarization step between each acquisition). The $850$ transient signals are coherently averaged for as long as the bosonic field remains phase-coherent. The coherence time of the bosonic fields is on the order of $10^6$ periods of oscillation, which for fields of $\sim 19$ Hz corresponds to $\sim 14$ hours. This limits the search bandwidth to fields of frequencies below $19$ Hz  (corresponding to bosons with mass lower than $ 7.8 \times 10^{-14}$ eV).
	
	\section{\textbf{Dark-matter-field directionality}}\label{s:dailyannualmodulation}
	\subsection{\textbf{Daily and annual modulations}}
	We note that the ALP field described in Eq. \eqref{axionfield}, seen in the Earth's reference frame, has annual and daily modulations due to the motion of the Earth in the Solar System. These modulations are expected to \textbf{1}- modulate the orientation of the experiment's sensitive axis with respect to the dark-matter wind, thus modulating the overall experimental sensitivity and \textbf{2}- change the lineshape of the experimental power spectra from pure Lorentzians to skewed Lorentzians. These effects are addressed and accounted for during the data processing by weighting transient spectra during the averaging sequence using the following method.
	
	For axions and ALPs, the experiment is sensitive to the projection of the bosonic pseudo-magnetic field gradient, $\bs{\nabla}a(\bs{r},t)$, onto the direction of the leading magnetic field. We denote this axis in the laboratory frame by the unit vector $\bs{\hat{e}}\ts{z}$. As a result of the yearly revolution of the Earth around the Sun and the daily revolution of the Earth, the alignment of $\bs{\hat{e}}\ts{z}$ with $\bs{\nabla}a(\bs{r},t)$ varies over time. This effect can increase or totally suppress the expected signal depending on when transient acquisitions happen and must be included in the analysis to avoid overestimating the experiment's sensitivity. 
	
	To this aim, we use the analysis proposed in Ref. \cite{Bandyopadhyay2012}, originally applied to nuclear recoil in WIMP-nucleus scattering but which directly translates to this sensitive-axis misalignment. Detailed and comprehensive calculations are given  in Ref. \cite{Bandyopadhyay2012}, here we show how these results are applied to our current signal analysis. We recall that the ALP scalar-field gradient can be written as (ignoring its initial phase):
	\begin{align}
	\bs{\nabla}a(\bs{r},t)=   a_0 \cos(\omega\ts{DM} t - \bs{k}\cdot\bs{r} - \pi/2) \cdot\bs{k}~,
	\end{align}\label{eq:grad}
	where $\omega\ts{DB} = \sqrt{(m_{\text{DM}} c^2)^2 + (m_{\text{DM}} vc)^2}/\hbar$ is the bosonic particle de Broglie frequency, $\bs{k}= m_{\text{DM}} \bs{v}/\hbar$ is its wave-vector with $\bs{v}$, the average velocity of the particles in the laboratory frame and $m_{\text{DM}}$ is the mass of the particle. When shifting from the galactic-center to the laboratory frame, $\bs{v}$ can be written as the following vectorial sum of velocities:
	\begin{align}\label{eq:mod}
	\bs{v}(t) = \bs{v}_{\text{DM}} - \bs{v}_{_{\text{SG}}} - \bs{v}_{_{\text{ES}}}(t) - \bs{v}_{_{\text{LE}}}(t)~,
	\end{align}
	where $\bs{v}_{\text{DM}}$ is the  velocity of the local bosonic field with respect to the center of the galaxy and is taken to be zero on average due to the isotropic structure of the dark-matter halo, $\bs{v}_{_{\text{SG}}}$ is the velocity of the Sun with respect to the center of the galaxy, i.e. $\bs{v}_{_{\text{SG}}}$ is the velocity of the Sun towards the Cygnus constellation and is approximatly constant,   $\bs{v}_{_{\text{ES}}}(t)$ is the velocity of the center of the Earth with respect to the sun and $\bs{v}_{_{\text{LE}}}(t)$ is the velocity of the laboratory with respect to center of the Earth. The expected signal amplitude $S,$ is proportional to the projection of the field's gradient onto the sensitive axis:
	\begin{align}
	S(t) &\propto \bs{\nabla}a(\bs{r},t)\cdot \bs{\hat{e}}\ts{z}\\
	&\propto  \frac{m_{\text{DM}}  a_0}{\hbar} \cos(\omega\ts{DB} t -\bs{k}\cdot\bs{r}) \bs{v}(t)\cdot\bs{\hat{e}}\ts{z}\\
	\label{eq:vtoz}	  &\propto  \frac{m_{\text{DM}}  a_0}{\hbar} \cos(\omega\ts{DB} t - \bs{k}\cdot\bs{r})...\\
	&\phantom{q} \times \Big(  \bs{v}_{_{\text{SG}}}\cdot\bs{\hat{e}}\ts{z} + \bs{v}_{_{\text{ES}}}(t)\cdot\bs{\hat{e}}\ts{z} + \bs{v}_{_{\text{LE}}}(t)\cdot\bs{\hat{e}}\ts{z}\Big)~.\nonumber 
	\end{align}
	
	During the calibration and benchmark acquisitions, an oscillating-magnetic field is continuously applied along the $\bs{\hat{e}}\ts{z}$ direction. After averaging the benchmark sideband amplitude $\text{A}_{\text{s}_0}$ is obtained from the fitting procedure. To avoid overestimating the expected signal amplitude and account for the previously discussed astrophysical motions, we weigh the transient spectra by the factor $\frac{|\bs{v}(t_n)\cdot\bs{\hat{e}}\ts{z}|}{|v|}\approx \frac{|\bs{v}(t_n)\cdot\bs{\hat{e}}\ts{z}|}{10^{-3}c}$ before including them in the average spectra, where $t_n$ is the starting time of the $n^{\text{th}}$ transient acquisition. This replicates the daily and yearly amplitude modulations of the signal induced by the motion of the Earth in the Solar System such as to avoid overestimating the expected signal amplitude from the bosonic fields.

	During the search acquisitions, no oscillating-magnetic field is applied and this misalignment effect is naturally included in the data. Nonetheless we weight the transient spectra by the factor $\frac{\bs{v}(t_n)\cdot\bs{\hat{e}}\ts{z}}{|v|}$ before averaging them.  Indeed, a transient spectrum  acquired when $\bs{v}(t_n)\cdot\bs{\hat{e}}\ts{z}=0$, cannot possibly include any signal and thus contains only noise. The effect of weighting such a transient spectrum by this factor is to prevent the addition of noise in the averaged spectrum. On the contrary, spectra acquired when $\bs{v}(t_n)\cdot\bs{\hat{e}}\ts{z}$ is maximal possess higher weight to account for the fact that the signal should be maximal during the acquisition.
	
	For dark photons, the experiment is sensitive to the projection of the field's polarization, denoted by the unit-vector $\hat{\bs{\epsilon}}$, onto the leading field direction $\bs{\hat{e}}\ts{z}$. We note that, at any point in space and time, $\hat{\bs{\epsilon}}$ can take any direction and is uniformly distributed along the unit-3D-sphere. Moreover, $\hat{\bs{\epsilon}}$ is assumed to be constant during one coherence time, which is always longer that the total measurement time. Therefore the experiment probes a unique value of $\hat{\bs{\epsilon}}$, modulated by the annual and daily motions of the sensitive axis of detection.

	The analysis is performed assigning three equally probable orientations $\hat{\bs{\epsilon}}_i$; and by time propagating $\hat{\bs{\epsilon}}_i\cdot \bs{\hat{e}}\ts{z}(t_0)$. This quantity is then used as a weight during the averaging of transient signal similar to the ALP case, thus yielding three corresponding limits. 
	
	To obtain limits comparable with other experiments, the three $\hat{\bs{\epsilon}}_i$, are given in the non-rotating Celestial Frame. The axis of this frame are practically fixed on any relevant time scale, since it takes the Solar System $\sim250$ million years to complete one orbit about the galactic centre, so different experiments can be compared easily in this case (even if they are performed many years apart in time).

	The three assigned directions are the usual orthonormal axis of the non-rotating Celestial Frame. The first assigned orientation, $\hat{\bs{\epsilon}_1}$, lies in the Celestial Equator Plane and points towards the Sun at the Vernal Equinox. The second orientation,   $\hat{\bs{\epsilon}_2}$, points towards the North Celestial Pole (i.e. along the axis of rotation of the Earth). In this case the sensitivity of the experiment is null, as $\bs{\hat{e}}\ts{z}$ is pointing East in the laboratory frame and remains perpendicular to  $\hat{\bs{\epsilon}_2}$ at any point in time. The third orientation is the following vector product $\hat{\bs{\epsilon}_3}=\hat{\bs{\epsilon}_1}\times\hat{\bs{\epsilon}_2}$, and lies in the Celestial Equator Plane.
	A complete mathematical description and schematic of this coordinate system can be found in Ref. \cite{Bandyopadhyay2012}.
	
	\subsection{\textbf{Search data time-stamp}}
	The search data are composed of $850$ transient scans, each $30$~s long, spaced by a polarization time of $31$~s. The first transient scan was acquired on the $28^{\text{th}}$~of~July~$2018$~at~$10$:$00$:$06$ (CET). Data acquired in Mainz, Germany.
	
	\subsection{\textbf{Expected sideband lineshape}}\label{s:lineshapre}
	The oscillating bosonic fields described in the Earth frame also exhibit annual and daily modulations due to the presence of the modulated wave-vector $\bs{k}$, and de Broglie frequency $\omega\ts{DB}$ in the arguments of the cosine function. The effect of such  modulations is to modify the lineshape of the power spectra from the expected Lorentzian signal. The first modulation due to the motion of the Earth with respect to the galactic center enters in the de Broglie frequency $\omega\ts{DB} = \sqrt{(m_{\text{DM}} c^2)^2 + (m_{\text{DM}} vc)^2}/\hbar$, where $\bs{v}$ is modulated as in Eq. \eqref{eq:mod}. The amplitude of the velocities in Eq. \eqref{eq:mod} are \mbox{$|\bs{v}_{\text{DM}}|\approx 0$, $|\bs{v}_{_{\text{SG}}}|\approx 220$ km.s$^{-1}\approx 10^{-3}c$}, $|\bs{v}_{_{\text{ES}}}(t)|\approx 30$ km$/$s and $|\bs{v}_{_{\text{LE}}}(t)|\approx 0.5$ km$/$s \cite{Bandyopadhyay2012}. Those components are therefore negligible compared to the rest mass-energy of the particle, so the usual approximation $\omega\ts{DB} \approx m_{\text{DM}} c^2/\hbar$ is used. The second modulation appears in the spatial argument of the cosine function, $\bs{k}\cdot\bs{r}$. This component is dominated by the velocity of the Sun towards the Cygnus constellation, $|\bs{v}_{_{\text{SG}}}|$, which is taken as constant both in amplitude and direction. Therefore, we neglect other components of the wave-vector. These two approximations yield a non-modulated cosine-form gradient, thus yielding a pure Lorentzian power-spectral line.
\end{document}